\documentclass
{aastex631}

\usepackage{CJK}
\usepackage{multirow}
\usepackage{graphicx}
\usepackage{float}    
\usepackage{orcidlink}

\begin{document}

\begin{CJK*}{UTF8}{gbsn}


\title{Solar Plasma Noise in TianQin Laser Propagation: An Extreme Case and Statistical Analysis
}

\correspondingauthor{Su, W.; Zhang, X. F.}
\email{suwei25@mail.sysu.edu.cn, zhangxf38@mail.sysu.edu.cn}


\author{Yanan Liu $\dagger$}
\author{Wei Su (苏威) $\dagger$ \orcidlink{0000-0002-6613-957X}}
\affiliation{TianQin Research Center for Gravitational Physics \& School of Physics and Astronomy, \\
Sun Yat-sen University (Zhuhai Campus), Zhuhai 519082, People’s Republic of China}
\affiliation{These authors contributed to the work equally.}


\author{Xuefeng Zhang}
\affiliation{TianQin Research Center for Gravitational Physics \& School of Physics and Astronomy, \\
Sun Yat-sen University (Zhuhai Campus), Zhuhai 519082, People’s Republic of China}

\author{Jixiang Zhang (张吉祥)}
\affiliation{TianQin Research Center for Gravitational Physics \& School of Physics and Astronomy, \\
Sun Yat-sen University (Zhuhai Campus), Zhuhai 519082, People’s Republic of China}

\author{Shenwei Zhou (周莘为)}
\affiliation{TianQin Research Center for Gravitational Physics \& School of Physics and Astronomy, \\
Sun Yat-sen University (Zhuhai Campus), Zhuhai 519082, People’s Republic of China}









\begin{abstract}

TianQin proposes to detect gravitational wave signals by using laser interferometry. 
However, laser propagation effect introduces a potential noise factor for TianQin.
In this work, we used MHD simulations to obtain the space magnetic field and plasma distributions during an extremely strong solar eruption,
and based on the magnetohydrodynamic simulation result, we investigate laser propagation noise for TianQin.
For the extremely strong solar eruption event, we find that the laser propagation noise closely approaches 100\% of TianQin's displacement noise requirement for Michelson combination; While the laser propagation noise is still about 30\% of TianQin's displacement noise requirement for time-delay interferometry X combination.
In addition, we investigate the laser propagation noise for 12 cases with different solar wind conditions. Our finding reveals a linear correlation between the laser propagation noise and several space weather parameters, e.g., solar wind dynamic pressure, Sym-H, and $D$st, where the correlation coefficients for solar wind dynamic pressure is strongest. 
Combining the cumulative distribution of solar wind dynamic pressure from 1999 to 2021 with the linear correlation between solar wind dynamic pressure and laser propagation noise, we have determined that the occurrence rate of the laser propagation noise to be greater than 30\% of TianQin's displacement noise requirement for Michelson combination over the entire solar activity week is about 15\%. In addition, we find that time-delay interferometry can suppress the laser propagation noise, and reduce the occurrence rate of the laser propagation noise exceeding 30\% of TianQin's requirement to less than 1\%.

\end{abstract}



\section{Introduction} \label{sec:intro}

Gravitational waves (GWs) play a crucial role in testing Einstein's general relativity theory and comprehending the mysteries of the universe's initial stages. Currently, there are two forms of gravitational wave detection programs: ground-based and space-based. The direct signal of the gravitational wave 
was firstly detected by LIGO \citep{Abbott2016}, a ground-based program. However, detecting gravitational waves below 10 Hz on the ground is an extreme challenge due to the limitations of ground vibration noise as well as gravitational gradient noise \citep{Amaro-Seoane2017, Luo2016TQ}. Therefore, in order to study the properties of gravitational waves and their sources over a wider range of frequencies, it is essential to implement a space-based program for the detection of gravitational waves. Currently, various projects are underway to detect gravitational waves in space, including Laser Interferometer Space Antenna (LISA) \citep{Amaro-Seoane2017}, DECIGO \citep{Kawamura2006DECIGO}, Taiji \citep{hu2017taiji}, and TianQin (TQ) \citep{Luo2016TQ}.

Among various programmes, DECIGO, developed in Japan, uses the interferometric mode of the Fabry-Perot resonance cavity to identify gravitational waves within the 0.1 -- 10 Hz frequency range. Meanwhile, TQ, LISA, and Taiji employ three drag-free satellites to constitute a constellation that approximated to an equilateral triangle. They detect gravitational-wave signals by using laser interferometry in the sensitive $10^{-4}$ -- 1 Hz frequency range. LISA and Taiji both adopt heliocentric orbits, the arm length of LISA and Taiji are 2.5 million km and 3 million km, respectively. Whereas, TQ adopts a geocentric orbit, with the satellite orbiting at an altitude of approximately 100000 km, the distance between each two satellites of TQ is about 170000 km, and the period of TQ is about 3.65 days \citep{Qiao2023, Jia2023, Ye2021}. For the orbit of TQ, the satellite and laser links will pass through the solar wind, bow shock, magnetopause, and magnetotail \citep{Su2020, Su2021}.
Compared to the heliocentric ones, TQ will be more significantly influenced by the interaction between solar wind and the magnetosphere.

The region surrounding the orbit of GWs detectors is not a vacuum but filled with medium, e.g., plasma, magnetic field, and energetic particles \citep{Chen2011}. 
There is acceleration noise caused by the interaction between space magnetic field and test mass \citep{Armano2020,Su2020, Su2023}, and the energetic particles (e.g., GCR, SEP) can cause continuous charging on test masses \citep{Lei2024,Han2023,Han2024,SunZhiyin2023}.
Besides, the dispersion effect during the propagation of the laser in space plasma can lead to optical path difference (OPD) noise for the laser interferometry range measure of GW detectors. This effect has been revealed and studied both by LISA and TQ \citep{Smetana2020, Jennrich2021, Su2021, Lu2021}. The laser propagation noise of a single laser link was investigated by magnetohydrodynamic (MHD) simulation for TQ \citep{Lu2021}, and analytical method for LISA \citep{Jennrich2021}, respectively. \citet{Su2021} revealed that time-delay interferometry (TDI) can suppress the laser propagation noise in time domain, and can suppress laser propagation noise in low frequency range. \citet{Jing2022J} used the PPMLR-MHD model with high spatial-temporal resolution combined with Kolmogorov's statistical theory of turbulence to analyse laser propagation noise.

However, 
our previous works have only investigated the laser propagation noise for TianQin (TQ) under moderate solar wind conditions \citep{Su2021,Lu2021,Jing2022J}.
We found that the optical path difference (OPD) noise due to laser propagation in space plasma can reach up to 30\% of the TQ’s requirement for the Michelson interferometric combination, which
suggests that laser propagation noise must be taken into account for TQ. So far it remains unknown whether solar eruptions lead to an increase in the laser propagation noise and if so, the magnitude of the increase.
Moreover, we do not have a statistical result for laser propagation noise during solar cycles yet.
As the period of solar activity cycle is about 11 years \citep{Hao2015}, and the solar wind parameters (e.g., magnetic field, solar wind speed, ion and electron number density) also has a period of variation of about 11 years \citep{Samsonov2019}. 
Due to the geocentric orbit of TQ, it may encounter magnetic storms \citep{MagneticStorms1964}, magnetic reconnections\citep{Oka2023,HuangSY2021,LiWY2021}, interplanetary shocks \citep{Kilpua2015}, coronal mass ejections \citep[CME;][]{ShenYD2012_1,ChenY_2007}. 
All these phenomena are related to the solar activity cycle \citep{Hathaway2015}, and there is a period of about 11 years for the occurrence rate of these eruption phenomena in heliosphere. 
This study aims to investigate the laser propagation noise with extreme solar activity condition, and attempts to obtain statistical results of the laser propagation noise for TQ.

This work is organised as follows: Section \ref{sec:Model and data} describes the space plasma model used in this study, along with the input data of the model, and the calculation of the laser propagation noise. Section \ref{sec:extremely strong case} analyses the laser propagation noise for an extremely intense case. Section \ref{sec:Statistical analysis} carries out a statistical analysis study of the laser propagation noise. Finally, Section \ref{sec:conclusion} presents the conclusions.


\section{Models and Data} \label{sec:Model and data}

\subsection{Laser Propagation Theory} \label{subsec:equation}

The solar wind is a plasma stream that originates from the solar atmosphere, and is composed mainly of electrons and protons \citep{pisacane2008space}. Solar wind permeates and affects the whole heliosphere. 
The solar wind interacts with the Earth's magnetic field to form the Earth's magnetosphere, and perturbations in the solar wind can also produce the phenomena such as magnetic storms, auroras, and so on \citep{Kallenrode2004}.
In the vicinity of Earth ($\approx 1$ AU), the typical temperature of the solar wind is about 150,000 K, with the typical solar wind speed of about 400 km/s, and the typical number density of about 1–10 particles per cm$^{-3}$ \citep{pisacane2008space}. 
As the laser travels through the space plasma, the variations of the electron number density can cause changes in the time delay of the electromagnetic (EM) wave propagation, resulting in the OPD noise that ultimately affects the accuracy of the displacement measurement.


The Appleton-Hartree (A-H) Equation can describe the refractive index $\mu$ for a laser with frequency $f$ (angular frequency $\omega$) travels through the Earth's magnetosphere and solar wind plasma \citep{2002ppd..book.....H}. Due to the space plasma around the laser link of TQ is collisionless, the electron collision frequency $\nu$ in the magnetosphere and solar wind plasma is much lower than the TQ laser frequency $\omega$, and the electron cyclotron frequency $\omega_{\rm B}$ being much lower than $\omega$, consequently, A-H Equation can be simplified as the following form (more details are in \citet{Su2021}),
\begin{equation}
\mu^2\ =1-\frac{\omega_\mathrm{p}^2\ }{\omega^2}
\end{equation}
where $\omega_{\rm p}$ is the plasma frequency, $\omega_{\rm p}^2=\frac{e^2N}{m\varepsilon_0}$, $e$ is the elementary charge, $N$ is the electron number density, $m$ is the electron mass, $\varepsilon_0$ is the vacuum electric permittivity.
Thus, the time ($\tau$) needed for the laser to traverse a distance $L$ in the space plasma is given as follow,
\begin{equation}
\tau=\int_{L}\frac{\mathrm{d}s}{\nu_g}=\int_{L}\frac{\mathrm{d}s}{c/\mu_g}=\int_{L}{\frac{\mathrm{d}s}{c}\left(\frac{\partial\left(\mu\omega\right)}{\partial\omega}\right)}=\int_{L}{\frac{\mathrm{d} s}{c}\left(\frac{1}{\sqrt{1-\frac{\omega_p^2}{\omega^2}}}\right)}
\label{eq2}
\end{equation}
since the plasma frequency $\omega_{\rm p}$ is much lower than $\omega$, Equation (\ref{eq2}) can be simplified as,
\begin{equation}
\tau=\int_{L}\frac{\mathrm{d}s}{c}\left(1+\frac{KN}{2f^2}\right)
\end{equation}
where $K=\frac{e^2}{4\pi^2m\varepsilon_0}=80.6$ $\rm m^3\ s^{-2}$. The OPD noise during the laser propagation in the space plasma can be calculated as:
\begin{equation}
\Delta l=c\Delta\tau=\frac{K}{2f^2}\int_{L}\ N\mathrm{\mathrm{d}}s .
\label{eq-dL}
\end{equation}

\subsection{Michelson and TDI} \label{subsec:Michelson and TDI}

In order to detect the minuscule displacement change caused by GWs, laser interferometry is used in GWs detection, 
i.e., the light produced by the laser is split into two beams by a splitter, then the two beams propagate along two different arms, and then, after being reflected by two mirrors, they are re-synthesised into a single beam at the beam splitter, which forms an interferometric signal in the photon detector \citep{muratore2021time}.
One possible method for detecting GWs in space is heterodyne interferometry, in which two lasers with different frequencies create an interfering signal on a photodetector \citep{muratore2021time}. 
The OPD noise generated during the propagation of the laser sent by spacecraft $i$ and received by satellite $j$ is denoted as $\Delta l_{ij}$.
For the Michelson interferometer centred on S1, the laser propagation noise for the two interferometer arms ( S1 - S2 and S1 - S3) are $\Delta l_{12}$, $\Delta l_{21}$, $\Delta l_{13}$, $\Delta l_{31}$, respectively. Since the propagation time of the laser between a pair of TQ satellites ($\approx$ 0.6 s) is much smaller than the time resolution of the simulation (60 s), we set $\Delta l_{ij}$=$\Delta l_{ji}$ in this work. Therefore, the OPD noise of the Michelson combination can be given by

\begin{equation}
\Delta l=2\left(\Delta l_{12}-\Delta l_{13}\right)
\end{equation}

Time-delay interferometry (TDI) is applied in the space-borne GWs detection to remove the laser phase noise which caused by the coupling between unequal arm lengths of satellites and the instabilities of the laser frequency. The principle of TDI is to create a virtual interferometer with equal arm lengths by linearly combining data and delays. These combinations are then added and subtracted to remove laser phase noise. Two of the typical examples of TDI combinations, TDI-$\alpha$ and TDI-$X$, are used for the analysis in this paper. $\ s_{ij}\left(t\right)$ is the phase noise associated with the $\Delta l_{ij}\left(t\right)$, $s_{ij}\left(t\right)=2\pi\Delta l_{ij}\left(t\right)/\lambda$, where $L_{ij}$ is the distance between satellites $i$ and $j$. Set the light speed $c$ = 1, the propagation time of the laser between satellite $i$ and satellite $j$ is $L_{ij}/c = L \approx 0.6$ s. The laser propagation noise for TDI-$\alpha$ combination ($s_\alpha$) and TDI-$X$ combination ($s_X$) can be written as \citep{Su2021},
\begin{equation}
\label{TDI Alpha noise}
s_\alpha=\left[s_{12}\left(t-\Delta t\right)-s_{12}\left(t\right)\right]\frac{2L}{\Delta t}-\left[s_{31}\left(t-\Delta t\right)-s_{31}\left(t\right)\right]\frac{2L}{\Delta t}
\end{equation}
\begin{equation}
\label{TDI X noise}
s_X=2s_\alpha
\end{equation}

For the Michelson combination, the equivalent strain noise ASD ($\sqrt{S_n^M}$) of TQ
is as follow \citep{2018CQGra..35i5008H},
\begin{equation}
\label{Michelson sensitivity}
S_n^M=S_n^x+S_n^a\left(1+\frac{{10}^{-4}\mathrm{Hz}}{f}\right),
\end{equation}
where $\sqrt{S_n^x}$ is the requirement of equivalent strain noise ASD of displacement noise, and $\sqrt{S_n^a}$ is the requirement of equivalent strain noise ASD caused by residual acceleration.
The equivalent strain noise requirement of TDI-$\alpha$  ($\sqrt{S_n^\alpha}$), and TDI-$X$ ($\sqrt{S_n^X}$) can be given as follows \citep{1999ApJ...527..814A},
\begin{equation}
\label{TDI Alpha sensitivity}
S_n^\alpha=\left[4\sin ^2 \left(\frac{3\pi f L}{c}\right)+24\sin ^2 \left(\frac{\pi f L}{c}\right)\right]S_n^a+6S_n^x
\end{equation}
\begin{equation}
\label{TDI X sensitivity}
S_n^X=\left[4\sin^2\left(\frac{4\pi f L}{c}\right)+32 \sin ^2\left(\frac{2\pi f L}{c}\right)\right]S_n^a+16\sin ^2\left(\frac{2\pi f L}{c}\right)S_n^x
\end{equation}

\subsection{SWMF and OMNI} \label{subsec:SWMF and OMNI}

The OPD noise due to laser propagation in space plasma is proportional to the integral of the electron number density along the laser link. Therefore, the study of laser propagation noise requires the spatial distribution of the plasma number density over the laser link, whereas the $in$-$situ$ measurements can only provide the electron number density at one point.
In this work, we use an MHD model, the Space Weather Modeling Framework \citep[SWMF;][]{2005JGRA..11012226T}, to obtain the spatial distribution of the plasma. The SWMF is a publicly available model developed by the University of Michigan. It is developed to simulate the space physical processes of the Sun, interplanetary medium, the Earth and other planetary space environments.
The model is accessible through NASA's Community Coordinated Modeling Center (CCMC). 
In this work, the global magnetosphere module with radiation belt environment, inner magnetosphere, and ionosphere conductance modules of the SWMF is used in the simulations. Within these modules, we get the spatial and temporal variation of the space plasma around the orbit of TQ.
The OMNI solar wind datasets combined from several spacecrafts' observations, e.g., Advanced Composition Explorer \citep[ACE;][]{ACE1998}, Deep Space Climate Observatory \citep[DSCOVR;][]{DSCOVR2012}, Wind \citep{WIND1997}.
In this work, the input data are from OMNI and the observation of ACE.
The outputs of the SWMF are plasma parameters, e.g., velocity, number density, magnetic fields.
The SWMF is mainly used in space physics and space weather research. The goal is to investigate the complex interactions and space weather phenomena among celestial bodies in the solar system.
The SWMF is widely used in space weather research \citep{Gombosi2021}, particularly in modelling of the interaction between the solar wind and the magnetosphere \citep{Samsonov2016}.

The OMNI is the dataset of solar wind, as the solar wind is the source of the solar--terrestrial space environment, 
the OMNI dataset provides important scientific value in the study of heliospheric physics and important applications in monitoring and forecasting of space weather.
The OMNI data is taken as the input for the SWMF model here. The OMNI dataset integrates observational solar wind parameters from 20 spacecraft, providing plasma and solar wind data to map the nose of the bow shock, as well as solar activity and geomagnetic indices, and energetic particle flux. Additionally, OMNI dataset includes a range of combination parameters, e.g., plasma beta, plasma flow pressure, and the Alfven Mach number \citep{2005JGRA..110.2104K}.
Currently, the OMNI database provides solar wind data with temporal resolutions of 1 min, 5 min, and 1 h. In this work, temporal resolution of the MHD simulations is set as 1 min, which matches the highest temporal resolution of the input solar wind data.

\section{Laser propagation noise of an extremely strong case} \label{sec:extremely strong case}

Solar activities (e.g., solar wind, CMEs, flares) are the driver of the space weather \citep{ShenYD2012_2, ZhengRuisheng2020}, strong solar eruptions (e.g., CMEs and shocks) can significantly affect the space environment of the Earth \citep{Chen_zhou2022}.
When a powerful solar eruption encounters the Earth, it may greatly alter the structure of the Earth's magnetosphere, and cause the exchange of matter and energy between the magnetosphere and the solar wind to become violent \citep{Wang_C2009}, leading to the occurrence of strong magnetic storm and the production of auroras \citep{Kallenrode2004,Lu_J_Y2013,Lu_J_Y2019}. One of the strongest CMEs and flares during the 23rd solar activity cycle is observed in October to November 2003 \citep{veselovsky2004solar, 2008ApJ...684.1448M}, it is known as Halloween events. 
During the Halloween events, large amounts of plasma and magnetic field ejected from the solar and encountered the Earth, and the Earth's magnetosphere was severely compressed by the plasma flow, and the density of the magnetosheath was several or even ten times denser than that of moderate solar wind conditions.
As the OPD noise is proportional to the integral of the electron number density along the laser beam, 
the OPD noise may be enhanced dramatically when a strong solar eruption encounters the Earth. 
In this work, we investigate laser propagation noise during an extremely strong solar eruption case (Halloween event).

\begin{figure}[ht!]
\plotone{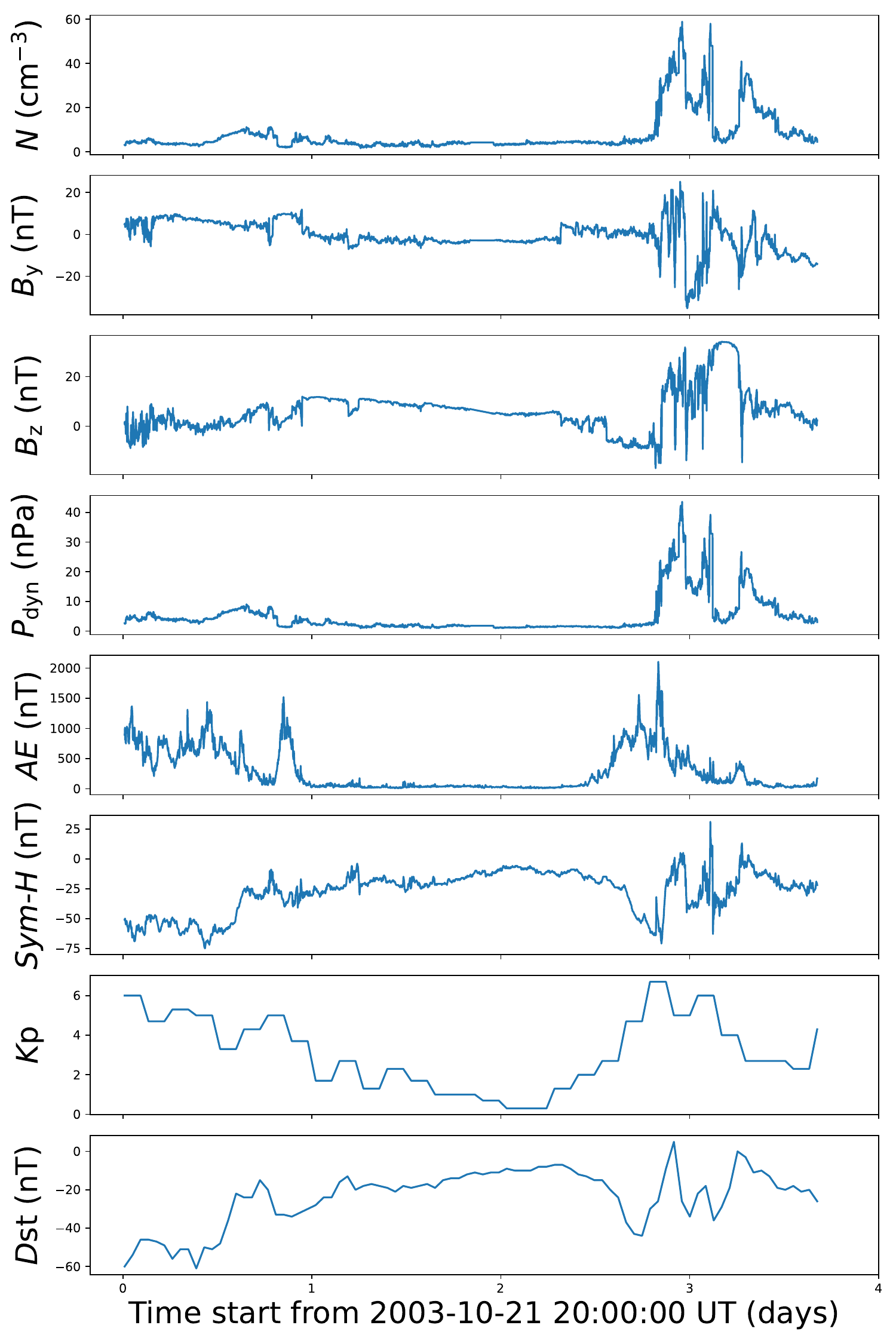}
\caption{The input data and space weather parameters during the extremely strong solar eruption.
From the top to bottom panels are electron number density, $y$ and $z$-component of the space magnetic field, $P_{\rm dyn}$, AE index, Sym-H index, Kp index, and $D$st index, respectively.
\label{fig: OMNI Data}}
\end{figure}

\begin{figure}[ht!]
\plotone{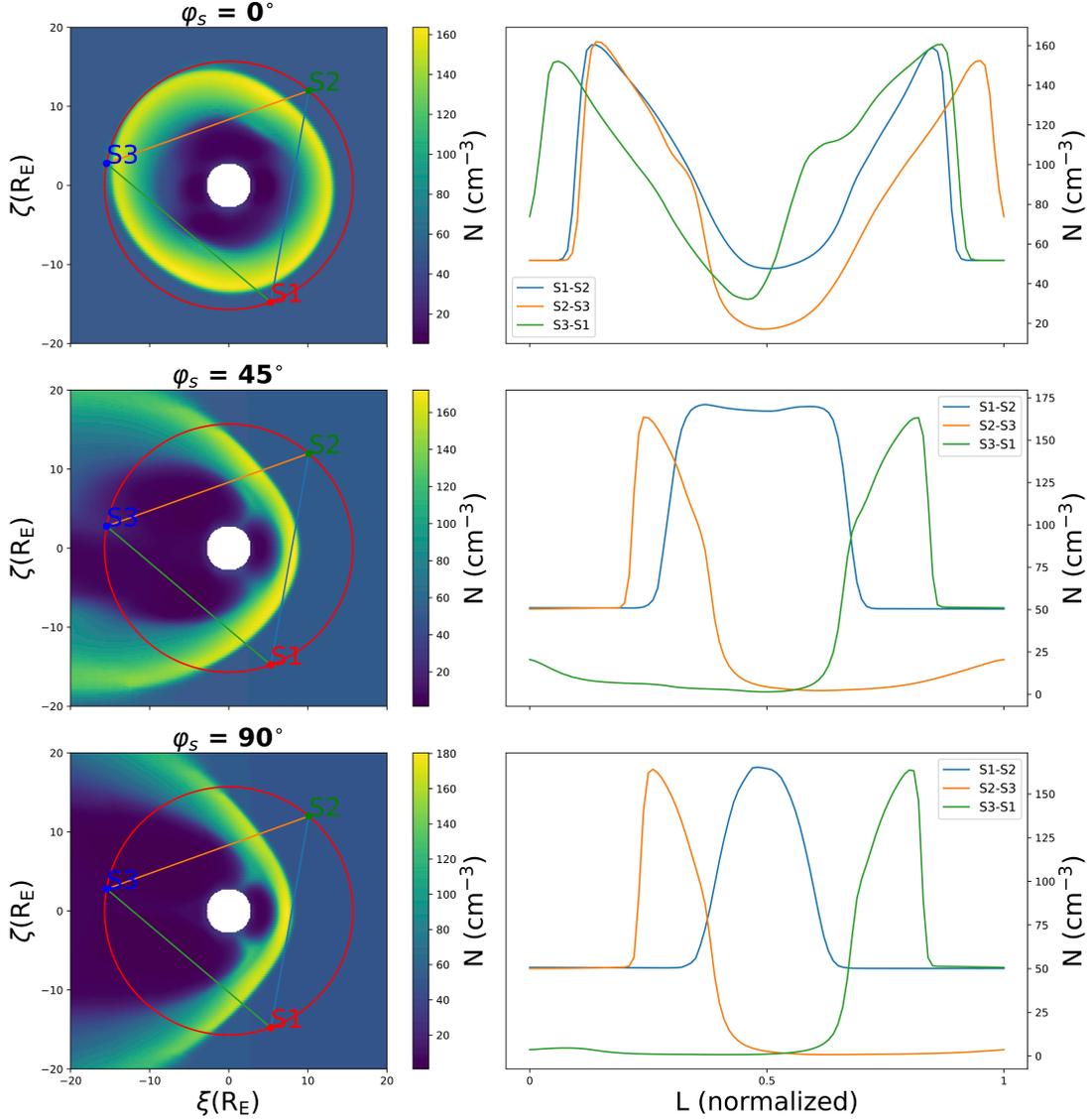}
\caption{The left column displays electron number density distribution on the orbital planes corresponding to $\varphi_s$=$0^\circ$, {45}$^\circ$ and {90}$^\circ$. The red circles represent the orbit of the TQ satellite, while the blue, orange, and green lines depict the S1-S2, S2-S3, and S3-S1 laser links, respectively. $\xi$ indicates the intersection line between the orbital plane and the ecliptic, and $\zeta$ is perpendicular to the ecliptic. The right column illustrates the electron number density distribution along three laser links, with colors corresponding to their counterparts in the left column.
\label{fig: 2D}}
\end{figure}

The input data and space weather parameters which are get from OMNI database during a Halloween event are shown in Figure \ref{fig: OMNI Data}. The input data (electron number density, the $y$ and $z$-component of the space magnetic field, and the solar wind dynamic pressure $P_{\rm dyn}$) are shown in the top 4 panels of Figure \ref{fig: OMNI Data} with the temporal resolution of 1 minute.
The space weather parameters AE and Sym-H indices are shown in the fifth and sixth panels of Figure \ref{fig: OMNI Data} with the temporal resolution of 1 minute.  
The space weather parameters Kp and $D$st indices are shown in the seventh and eighth panels of Figure \ref{fig: OMNI Data} with the temporal resolution of 1 hour.
With the input data of the MHD simulation, the SWMF is applied here to get the spatial and temporal distribution of space plasma around the TQ's orbit, and the results are shown in Figure \ref{fig: 2D}.
The left panels of Figure \ref{fig: 2D} show electron number density distribution maps at different orbital planes ($\varphi_s$=$0^\circ$, {45}$^\circ$ and {90}$^\circ$), where $\varphi_s$ is the angle between the normal direction of the TQ's orbital plane and the Sun--Earth line. 
The orbit of TQ is shown as red circles, while the laser links between each of the three satellites (denoted as S1, S2, and S3),  S1--S2, S2--S3 and S3--S1 are shown as blue, orange and green lines respectively. $\xi$ denotes the intersection of the orbital plane with the ecliptic, and $\zeta$ is perpendicular to the ecliptic. The electron number density distributions along these three laser links (S1--S2, S2--S3 and S3--S1) are shown in the right panels of Figure \ref{fig: 2D}, they are described as blue, orange and green, respectively, which are consistent with the colors of the laser links on the left panels. The maximum of the electron number density along the laser links corresponds to the magnetosheath and approaches 200 cm$^{-3}$, which is about 5 times of the moderate solar wind conditions in \citet{Su2021}.

\begin{figure}[ht!]
\plotone{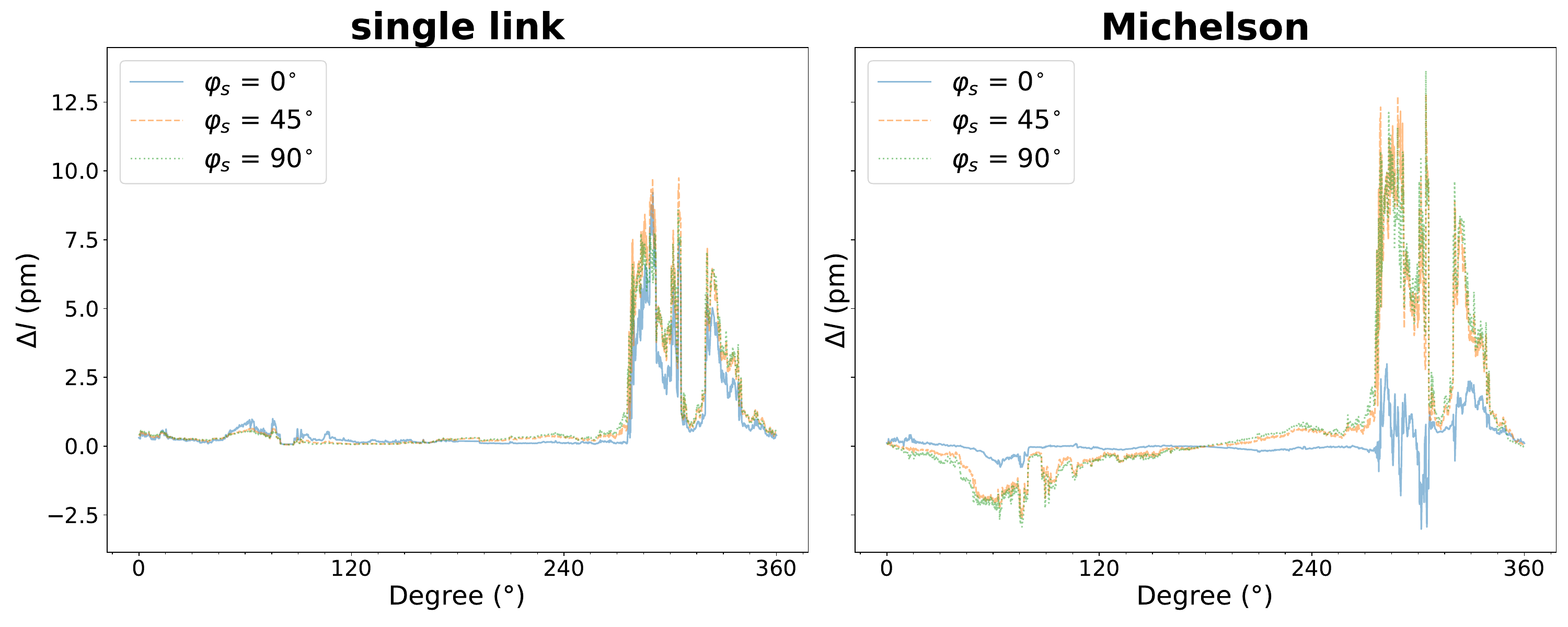}
\caption{The left panel is the OPD noise of the single (S1--S2) link, and the right panel is the OPD noise of the Michelson combination. The OPD noise of the constellation plane $\varphi_s=0^\circ$, 45$^\circ$, and 90$^\circ$ in a TQ orbit period (3.65 days) are represented as blue solid line, orange dashed line, and green dotted line, respectively.
\label{fig: singlelink-Michelson time}}
\end{figure}

Based on the distribution of electron number density along the laser links, we calculate the OPD noise by Equation (\ref{eq-dL}).
The left panels of Figure \ref{fig: singlelink-Michelson time} show the OPD noise of a single (S1--S2) laser link in one orbital cycle of TQ for three different orbital plane cases with $\varphi_s=0^\circ$(blue solid line), {45}$^\circ$ (orange dashed line), and {90}$^\circ$ (green dotted line). There is a little difference for the distribution of the OPD noise for the single laser link in the three orbital cases ($\varphi_s=0^\circ$, $45^\circ$, and $90^\circ$). The maximums of the OPD noise of the single laser link are nearly 10 pm for the extremely strong event, which is about one order larger than that of moderate solar wind conditions in \citet{Su2021}.

\begin{figure}[ht!]
\plotone{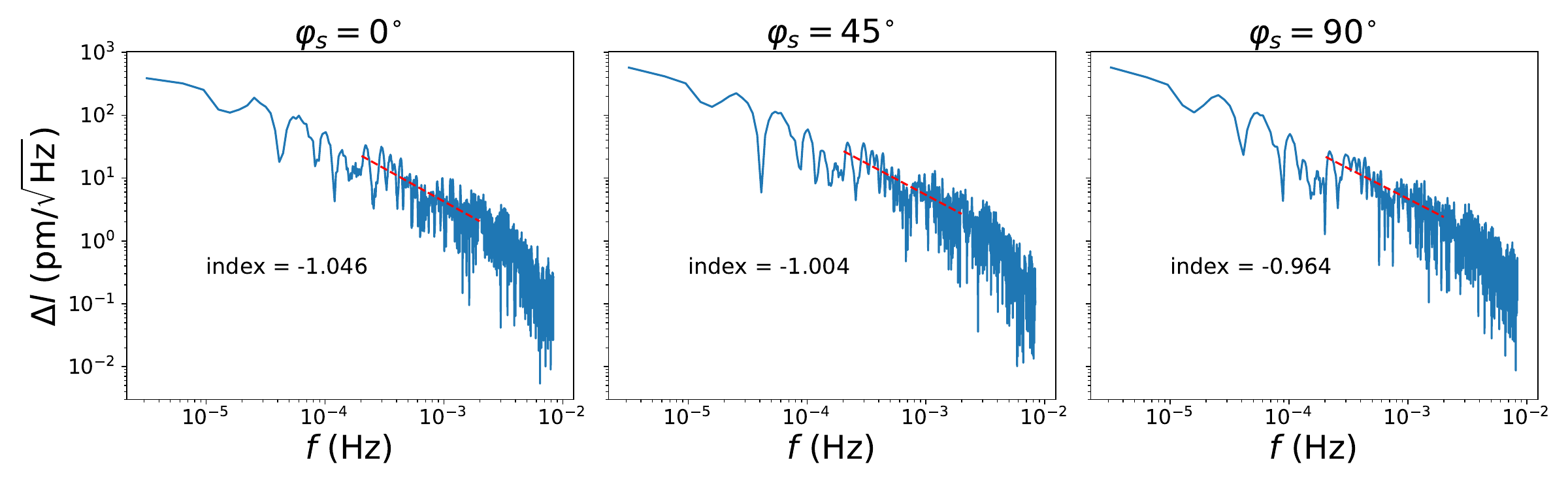}
\caption{The ASDs and power-law fitting results of OPD noise on the S1--S2 link in the constellation plane $\varphi_s=0^\circ$ (left panel), 45$^\circ$ (middle panel), and 90$^\circ$ (right panel) cases.
\label{fig: singlelink ASD}}
\end{figure}

Furthermore, we calculated the amplitude spectral distribution (ASD) of the OPD noise for single link, and the results are shown in Figure \ref{fig: singlelink ASD}. The red dashed line in Figure \ref{fig: singlelink ASD} is the fitted results of the ASD of the OPD noise. The spectral indices of the fitted results for the constellation planes $ \varphi_s=0^\circ$, {45}$^\circ$ and {90}$^\circ$ are -1.046, -1.004 and -0.964 respectively, and the amplitudes of the fitted results at 1 mHz are 4.252, 5.401 and 4.678 pm, respectively.

\begin{figure}[ht!]
\plotone{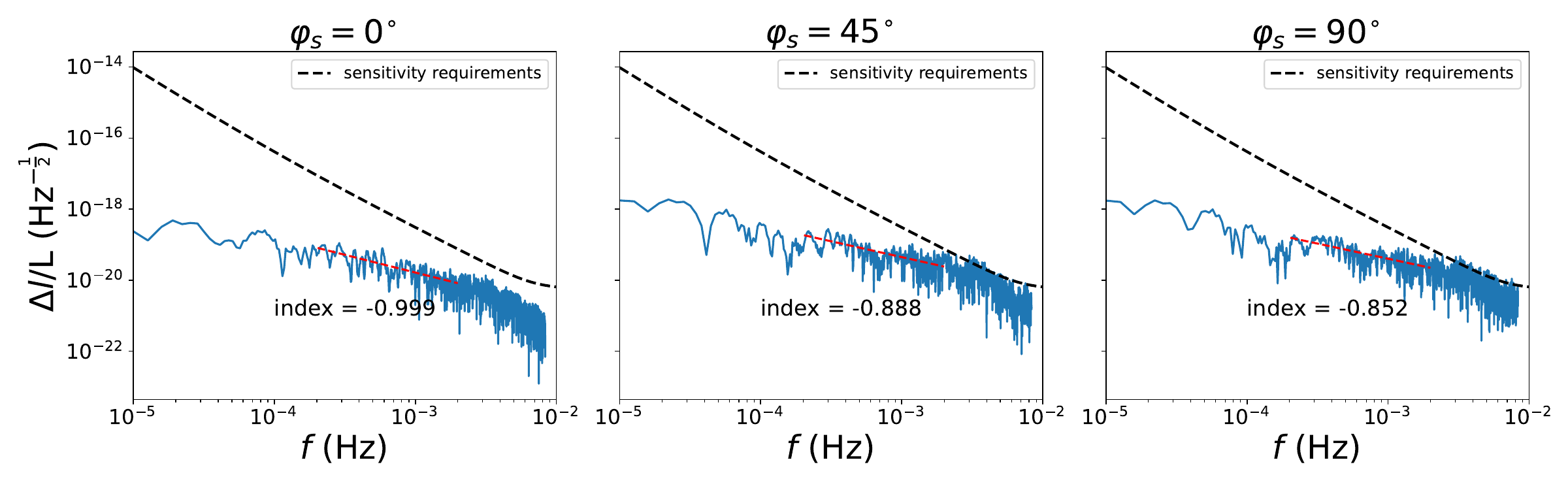}
\caption{The ASDs of the $\Delta l/L$ and the equivalent strain noise requirements of TQ for the Michelson combination in the constellation planes $\varphi_s=0^\circ$ (left panel), 45$^\circ$ (centre panel), and 90$^\circ$ (right panel), respectively.
\label{fig: Michelson ASD Sensitivity}}
\end{figure}

In the real GW detection mission, laser interferometry will be used to measure the changes of the distance between the test masses to obtain the properties of GWs signals, such as frequency, amplitude and polarisation.
The Michelson interference combination is the most typical laser interference combination. The right panel of Figure \ref{fig: singlelink-Michelson time} shows the time domain distribution of the OPD noise for the Michelson combination. The OPD noise in the constellation planes of $\varphi_s$=$45^\circ$ and $ \varphi_s$=$90^\circ$ is stronger than that of the single-link, while OPD noise of $\varphi_s$=$0^\circ$ case is weaker that of the single link. The OPD noise of $ \varphi_s$=$45^\circ$ and $ \varphi_s$=$90^\circ$ constellation planes is higher than the result of $\varphi_s$=$0^\circ$. 
Here, the equivalent strains of the OPD noise is denoted as $\Delta l/L$, where $L$ is the arm length of TQ. The results of $\Delta l/L$ of the Michelson combination in frequency domain are shown in Figure \ref{fig: Michelson ASD Sensitivity}.
The black dashed lines represent the equivalent strain noise requirement of the Michelson combination for TQ. The red dashed lines represent the fitted results of the ASD of the OPD noise. The fitted spectral indices of the constellation planes $\varphi_s$=$0^\circ$, ${45}^\circ$ and ${90}^\circ$ are -0.999, -0.888, and -0.852, respectively, with corresponding fitted spectral amplitudes at 1 mHz of 2.844 pm, 7.739 pm, and 6.970 pm, respectively. For the extremely strong solar eruption case, the spectral indices of the OPD noise are similar to that of the moderate solar wind conditions, whereas, the spectral amplitudes at 1 mHz are approximately one order larger than that of the moderate solar wind conditions in \citet{Su2021}.

\begin{figure}[ht!]
\plotone{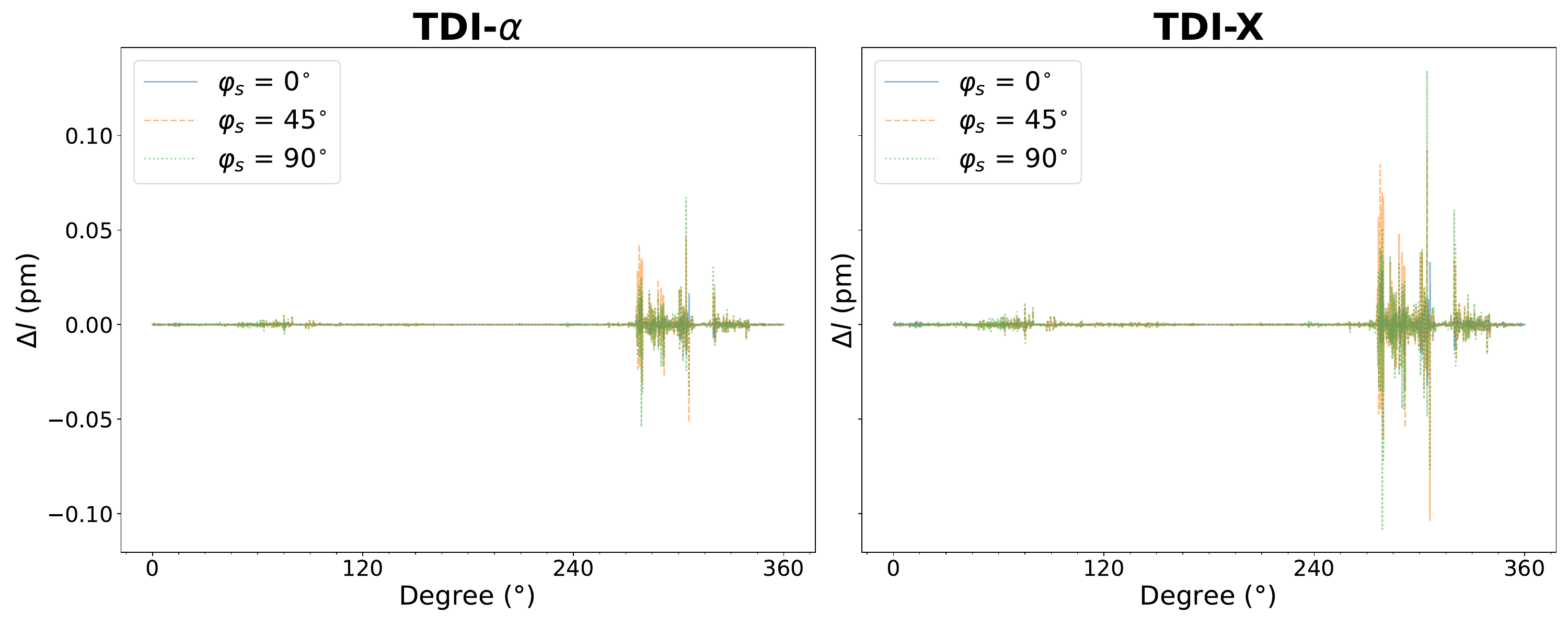}
\caption{The OPD noise for TDI-$\alpha$ combination (left) and TDI-X combination (right) in time domain, considering in the constellation planes of $\varphi_s=0^\circ$(blue solid line), 45$^\circ$ (orange dashed line), and 90$^\circ$ (green dotted line).
\label{fig: TDI time}}
\end{figure}

TDI is used in processing space-borne GWs detection data, and it is revealed that TDI can suppress the OPD noise \citep{Su2021}. Combining the electron number density along the laser links and OPD noise of TDI combinations (Equations (\ref{TDI Alpha noise}) and (\ref{TDI X noise})), we calculate the OPD noise for TDI-$\alpha$ and TDI-$X$ combinations. 
Figure \ref{fig: TDI time} shows the time domain distribution of the OPD noise for the combinations of TDI-$\alpha$ and TDI-$X$. The maximum absolute values of the OPD noise for the TDI-$\alpha$ and TDI-$X$ combinations noise are 0.067 pm and 0.134 pm, respectively. 
It suggests that the TDI combinations can lead to a 2 orders of magnitude decrease in OPD noise in the time domain compared to the Michelson combination.

\begin{figure}[ht!]
\plotone{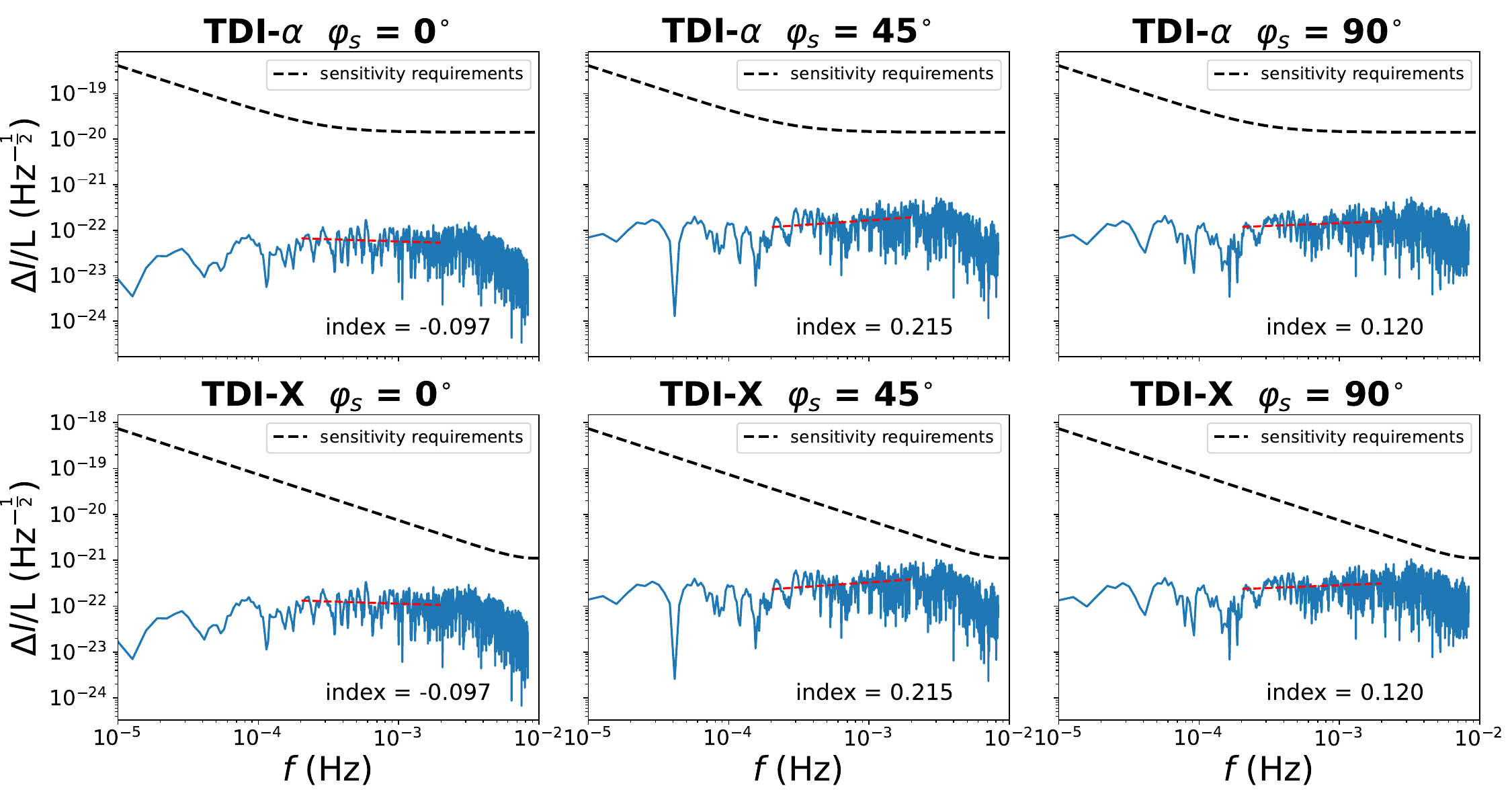}
\caption{The $\Delta l/L$ of TDI-$\alpha$ combination (top panels) and TDI-$X$ combination (bottom panels) for TQ in the case of the constellation planes $\varphi_s=0^\circ$ (left panels), 45$^\circ$ (centre panels) and 90$^\circ$ (right panels). The black dashed lines in the top and bottom panels are the equivalent strain noise requirements of TDI-$\alpha$ and TDI-$X$ combinations for TQ, respectively.
\label{fig: TDI ASD Sensitivity}}
\end{figure}

Furthermore, we calculate the ASDs of the OPD noise for TDI combinations.
Figure \ref{fig: TDI ASD Sensitivity} shows the
the equivalent strains of the OPD noise ($\Delta l/L$) for TDI-$\alpha$ and TDI-$X$ combinations.
The black dashed lines represent the equivalent strain noise requirements of TDI-$\alpha$ combination (top panels) and TDI-$X$ combination (bottom panels) for TQ.
The red dashed line shows the power law fitting results of the noise ASD in the frequency range. For the TDI-$\alpha$ combination (top row), the fitted indices are -0.0974, 0.2150 and 0.1200, respectively, with corresponding fitted amplitudes at 1 mHz of 0.0098, 0.0285, and 0.0247, respectively. Similarly, for the TDI-X combination (second row), the fitted indices are also -0.0974, 0.2150 and 0.1200, respectively, with corresponding fitted amplitudes at 1 mHz of 0.0196, 0.0570, and 0.0570, respectively.
It can be seen that the OPD noise in the low-frequency band ($f \lesssim $ 3 mHz) is significantly suppressed by the TDI combinations, and the OPD noise after the TDI combination is changed from colored noise to white noise in the low-frequency band ($f \lesssim $ 3 mHz).

\section{Statistical analysis of laser propagation noise} \label{sec:Statistical analysis}

The OPD noise is proportional to the integral of the electron number density along the laser link. 
On the spatial scale of the arms of TQ, about 170,000 km, the laser link will pass through magnetosphere, magnetosheath, bow shocks, solar wind, and so on. Among these regions, magnetosheath is the region of the highest number density along the laser link of TQ.
Various phenomena, e.g., fast and slow solar wind, magnetic storms, CMEs, and interplanetary shocks, can vary the geometric structure and the electron number density of these regions in multiple scales, from plasma dynamic scales in seconds to solar cycles in tens of years \citep{Sharma2005,Zhou2018}. Several physical parameters found to be sensitive to the geometric structure and the electron number density distributions, e.g., $y$- and $z$-component of the space magnetic field, solar wind speed, solar wind dynamic pressure ($P_{\rm dyn}$) \citep{Maezawa1976,Rufenach1992,Hartley2014,Wang_M2015}; 
And several space weather parameters can be used to characterise the structure of the Earth's magnetosphere, e.g., AE, Sym-H, Kp, and $D$st indices in Figure \ref{fig: OMNI Data}.
In general, with the increase of $P_{\rm dyn}$, the compression of the Earth's magnetosphere will increase, and so does the electron number density of the magnetic sheath.



\begin{figure}[ht!]
\plotone{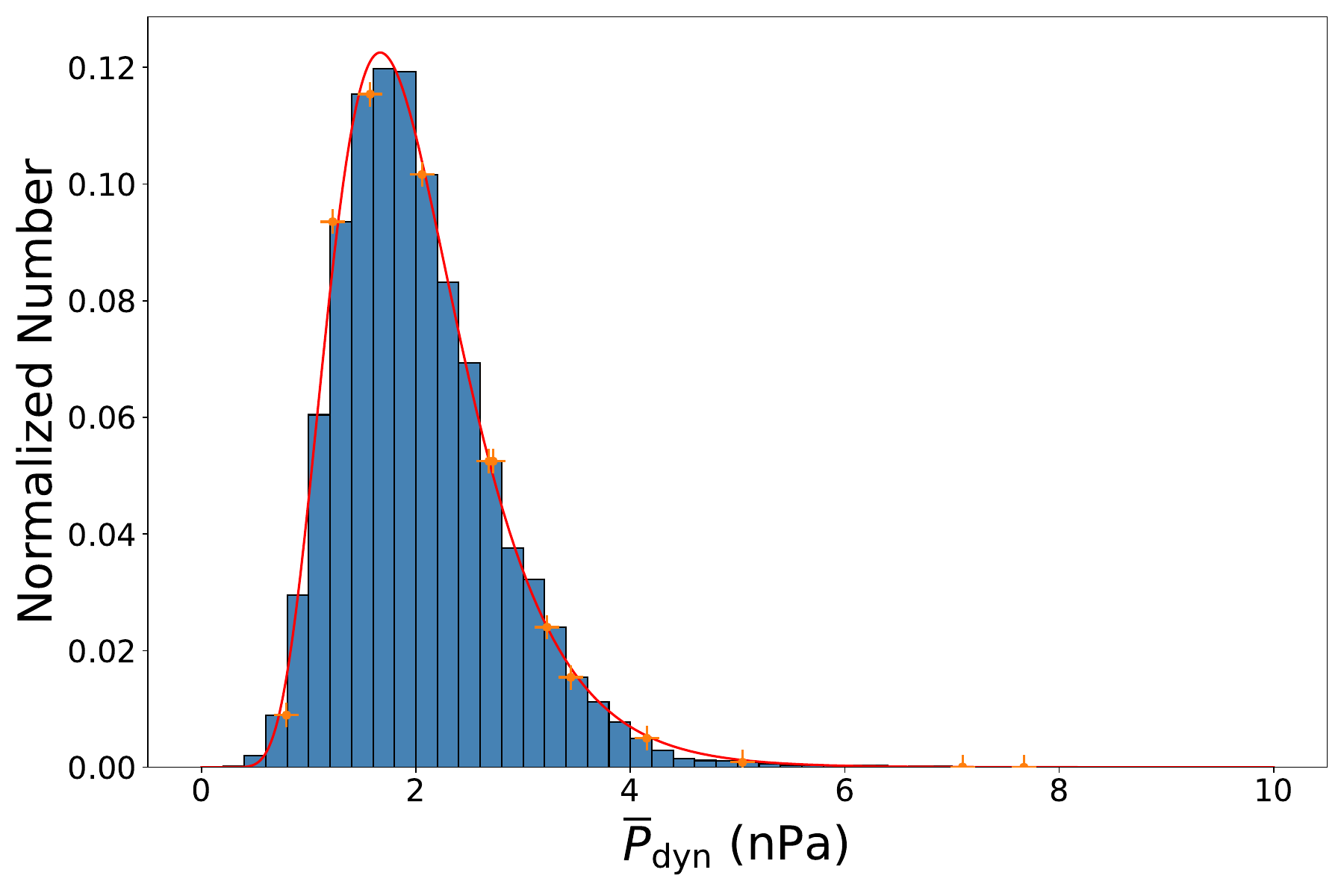}
\caption{Histogram of $\overline{P}_{\rm dyn}$
during 1999 to 2021. The blue bins are the histogram of $\overline{P}_{\rm dyn}$. The 12 orange pluses are the $\overline{P}_{\rm dyn}$ of the twelve cases in this work. The red curve is the fitted result of the histogram of $\overline{P}_{\rm dyn}$.}

\label{fig:hist-Pdyn}
\end{figure}

The MHD simulations of the space environment in the vicinity of TQ's orbit consume considerable computing resources. Even with parallel computing, it takes more than a month to simulate the space plasma data for one TQ orbital cycle (3.65 days). 
It would take more than 100 years to study the laser propagation noise over the entire solar activity cycle ($\sim$ 11 years).
Due to limited computational resources, we studied 12 cases of the OPD noise by using the SWMF with different input solar wind conditions.
In order to evaluate the relationship between the OPD noise and space weather parameters during different TQ orbital cycles, we define the average value of $P_{\rm dyn}$, $B_{\rm y}$, $B_{\rm z}$, AE, Sym-H, Kp, and $D$st during a TQ orbital cycle (3.65 days) as $\overline{P}_{\rm dyn}$, $\overline{B}_{\rm y}$, $\overline{B}_{\rm z}$, $\overline{\rm AE}$, $\overline{\rm Sym\text{-}H}$, $\overline{\rm Kp}$, and $\overline{D\rm st}$, respectively.
As shown in Figure \ref{fig:hist-Pdyn}, the blue bins are the histogram of $\overline{P}_{\rm dyn}$ during 1999--2021 (about 2 solar cycles), with the minimum and maximum of $\overline{P}_{\rm dyn}$ of 0.286 and 9.396 nPa, respectively.
In this work, we roughly selected twelve cases with $\overline{P}_{\rm dyn}$ of 0.79, 1.22, 1.57, 2.06, 2.68, 2.72, 3.22, 3.45, 4.15, 5.05, 7.10, and 7.67 nPa,
which are marked as orange pluses circles in Figure \ref{fig:hist-Pdyn}, and most range of $\overline{P}_{\rm dyn}$ during about 2 solar cycles are covered by the selected cases.

\begin{figure}[ht!]
\plotone{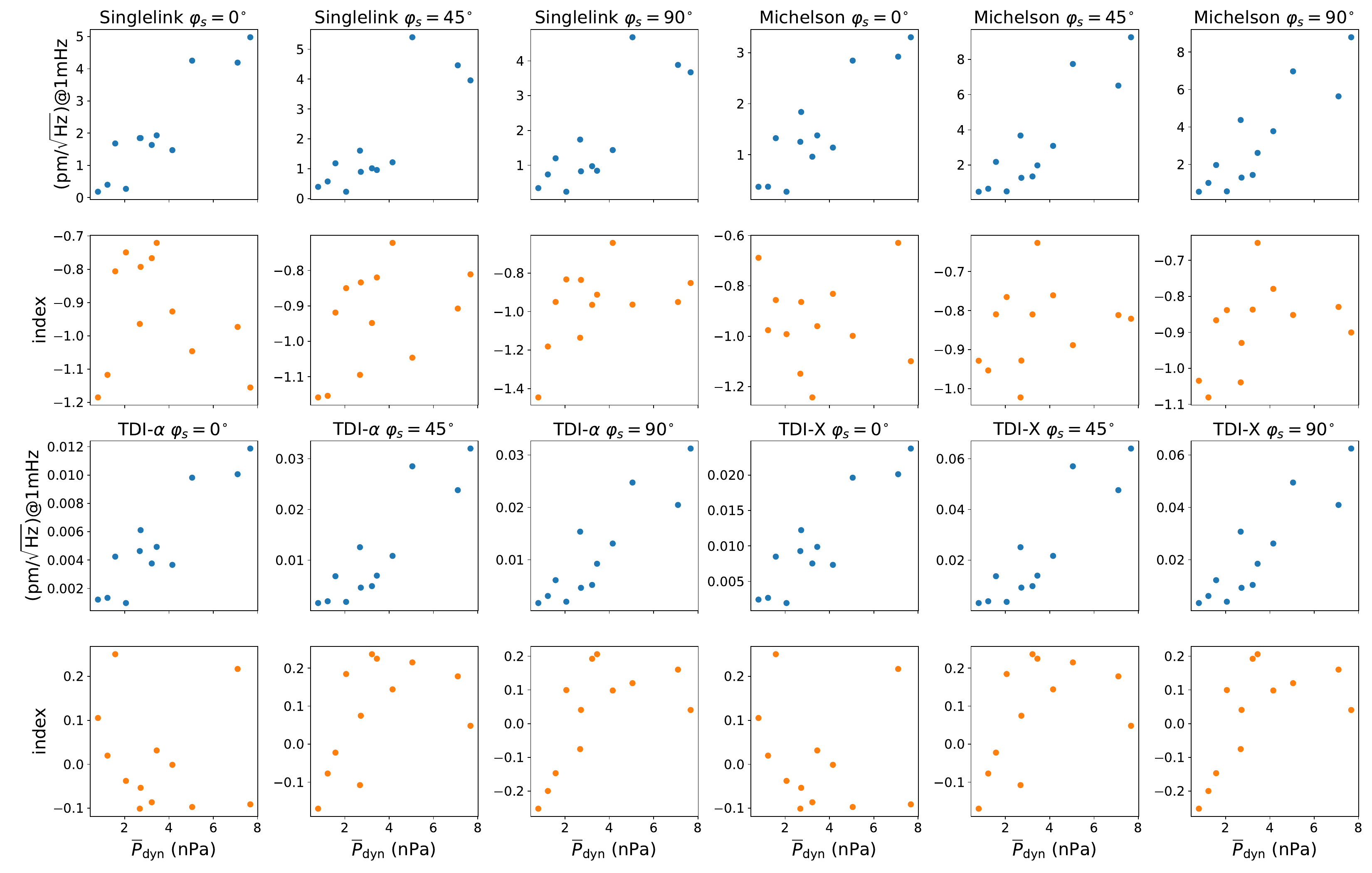}
\caption{Relationship between $\overline{P}_{\rm dyn}$ and power-law fitting of ASDs of the OPD noise of single link, Michelson, TDI-$\alpha$ and the TDI-X combinations for $\varphi_s$=$0^\circ$, {45}$^\circ$, and {90}$^\circ$, respectively. The blue scatters indicate the amplitude of OPD noise at 1 mHz, and the orange points are the fitted spectral indices of the ASDs of the OPD noise.
\label{fig: all fit}}
\end{figure}

Similar to the OPD noise study in Section \ref{sec:extremely strong case}, we use the SWMF to simulate the variation of the magnetosphere and obtain the electron number density distributions around the laser links of TQ, and then calculate the OPD noise for 12 TQ orbital cycles. 
In order to study the relationship between the OPD noise and space weather, we calculate the Pearson correlation coefficients between the power law fitted parameters (amplitude at 1 mHz and the) of the OPD noise's ASDs and space weather parameters ($\overline{P}_{\rm dyn}$, $\overline{B}_{\rm y}$, $\overline{B}_{\rm z}$, $\overline{\rm AE}$, $\overline{\rm Sym\text{-}H}$, $\overline{\rm Kp}$, and $\overline{D\rm st}$) for single link Michelson combination, TDI-$\alpha$ combination, and TDI-$X$ combination.
Figure \ref{fig: all fit} shows the relationship between the fitted of the ASD of the OPD noise at 1 mHz and $\overline{P}_{\rm dyn}$. Figure \ref{fig: all fit} contains results of $\varphi_s$=$0^\circ$, {45}$^\circ$ and {90}$^\circ$ constellation planes for the single link (top left 6 panels), the Michelson combination (top right 6 panels), the TDI-$\alpha$ combination (bottom left 6 panels), and the TDI-X combination (bottom right 6 panels). 
In Figure \ref{fig: all fit}, the horizontal axes are $\overline{P}_{\rm dyn}$. In the first and third rows, the vertical axes are the amplitude for the fitted ASDs of the OPD noise at 1 mHz, and the values are marked as blue dots. In the second and forth rows, the vertical axes are the fitted spectral indices for the ASDs of OPD noise, and the values are marked as orange dots.
There is an obvious linear correlation between the amplitude of the OPD noise and $\overline{P}_{\rm dyn}$;
While the correlation between the spectral index and $\overline{P}_{\rm dyn}$ is not apparent. 

Furthermore, we investigate the relationship between the OPD noise and space weather parameters quantitatively. We calculate the linear correlation coefficients between the power law fitted parameters (amplitude at 1 mHz and the power law indices) of the OPD's ASDs and space weather parameters ($\overline{P}_{\rm dyn}$, $B_{\rm y}$, $B_{\rm z}$, AE index, Sym-H index, Kp index, and $D$st index) for the single link, Michelson combination, TDI-$\alpha$ combination, and TDI-$X$ combination, and the results are shown in Table \ref{tab:cc}.
From the table we can find that most absolute of the linear correlation coefficients between the spectral indices and all seven space weather parameters are lower than 0.7, it indicates that there is no apparent linear correlation between the spectral indices and space weather parameters.
The absolute of the Pearson correlation coefficients between the amplitude of OPD's ASDs at 1 mHz and space weather parameters $\overline{P}_{\rm dyn}$, $\overline{B}_{\rm y}$, $\overline{B}_{\rm z}$, $\overline{\rm AE}$, $\overline{\rm Sym\text{-}H}$, $\overline{\rm Kp}$, and $\overline{D\rm st}$ are about 0.89, 0.27, 0.36, 0.58, 0.72, 0.69, and 0.78, respectively.
The above results show that for $\overline{B}_{\rm y}$ and $\overline{B}_{\rm z}$, there is no correlation between them and the amplitude of OPD's ASDs; For AE and Kp indices, there is a weak correlation between them and the amplitude of OPD's ASDs; For $\overline{P}_{\rm dyn}$, $\overline{\rm Sym\text{-}H}$, and $\overline{D\rm st}$, there is a significant linear correlation between them and the amplitude of OPD's ASDs, with the strongest linear correlation between $\overline{P}_{\rm dyn}$ and the amplitude of OPD's ASDs. 


\begin{table}[]
\caption{The correlation coefficients between the power law fitted parameters (amplitude at 1 mHz and the power law indices) of the OPD noise's ASDs and space weather parameters ($\overline{P}_{\rm dyn}$, $\overline{B}_{\rm y}$, $\overline{B}_{\rm z}$, $\overline{\rm AE}$, $\overline{\rm Sym\text{-}H}$, $\overline{\rm Kp}$, and $\overline{\mathit{D}\text{st}}$) for single link Michelson combination, TDI-$\alpha$ combination, and TDI-$X$ combination.}
\label{tab:cc}
\resizebox{\textwidth}{!}{$
\begin{tabular}{l|l|lll|lll|lll|lll}
\hline
Correlation Coefficient            & Parameter     & \multicolumn{3}{c|}{Single} & \multicolumn{3}{c|}{Michelson} & \multicolumn{3}{c|}{TDI $\alpha$} & \multicolumn{3}{c}{TDI $X$} \\ \hline
$\varphi_s$ ($^\circ$)                             &               & 0       & 45      & 90      & 0        & 45       & 90       & 0         & 45       & 90       & 0       & 45     & 90     \\ \hline
\multirow{7}{*}{Amplitude @ 1 mHz} & $\overline{P}_{\rm dyn}$ & 0.921   & 0.830   & 0.839   & 0.895    & 0.895    & 0.885    & 0.905     & 0.899    & 0.895    & 0.905   & 0.899  & 0.895  \\
                                   & $\overline{B}_{\rm y}$         & 0.243   & 0.210   & 0.259   & 0.179    & 0.277    & 0.367    & 0.184     & 0.278    & 0.365    & 0.184   & 0.278  & 0.365  \\
                                   & $\overline{B}_{\rm z}$         & 0.323   & 0.534   & 0.505   & 0.309    & 0.334    & 0.319    & 0.294     & 0.368    & 0.320    & 0.294   & 0.368  & 0.320  \\
                                   & $\overline{\rm AE}$            & 0.561   & 0.563   & 0.586   & 0.499    & 0.611    & 0.633    & 0.506     & 0.624    & 0.643    & 0.506   & 0.624  & 0.643  \\
                                   & $\overline{\rm Sym\text{-}H}$         & -0.729  & -0.689  & -0.684  & -0.654   & -0.769   & -0.739   & -0.680    & -0.769   & -0.746   & -0.680  & -0.769 & -0.746 \\
                                   & $\overline{\rm Kp}$            & 0.682   & 0.668   & 0.679   & 0.635    & 0.712    & 0.723    & 0.645     & 0.727    & 0.737    & 0.645   & 0.727  & 0.737  \\
                                   & $\overline{\mathit{D}\text{st}}$  & -0.777  & -0.733  & -0.731  & -0.718   & -0.822   & -0.815   & -0.736    & -0.825   & -0.824   & -0.736  & -0.825 & -0.824 \\ \hline
\multirow{7}{*}{Index}             & $\overline{P}_{\rm dyn}$ & -0.198  & 0.455   & 0.476   & -0.013   & 0.268    & 0.390    & -0.123    & 0.484    & 0.598    & -0.123  & 0.484  & 0.598  \\
                                   & $\overline{B}_{\rm y}$         & 0.151   & 0.403   & 0.540   & -0.141   & 0.399    & 0.463    & -0.088    & 0.425    & 0.489    & -0.088  & 0.425  & 0.489  \\
                                   & $\overline{B}_{\rm z}$         & -0.081  & -0.173  & 0.030   & 0.049    & 0.052    & 0.271    & -0.175    & 0.393    & 0.299    & -0.175  & 0.393  & 0.299  \\
                                   & $\overline{\rm AE}$        & 0.003   & 0.511   & 0.586   & -0.052   & 0.352    & 0.496    & -0.192    & 0.591    & 0.677    & -0.192  & 0.591  & 0.677  \\
                                   & $\overline{\rm Sym\text{-}H}$  & 0.310   & -0.287  & -0.288  & 0.223    & -0.333   & -0.319   & 0.260     & -0.446   & -0.454   & 0.260   & -0.446 & -0.454 \\
                                   & $\overline{\rm Kp}$      & -0.075  & 0.505   & 0.550   & -0.024   & 0.338    & 0.490    & -0.259    & 0.612    & 0.717    & -0.259  & 0.612  & 0.717  \\
                                   & $\overline{\mathit{D}\text{st}}$           & 0.274   & -0.401  & -0.357  & 0.016    & -0.426   & -0.448   & 0.140     & -0.468   & -0.539   & 0.140   & -0.468 & -0.539 \\ \hline
\end{tabular}
    $$
$}

\end{table}

\begin{figure}[ht!]
\plotone{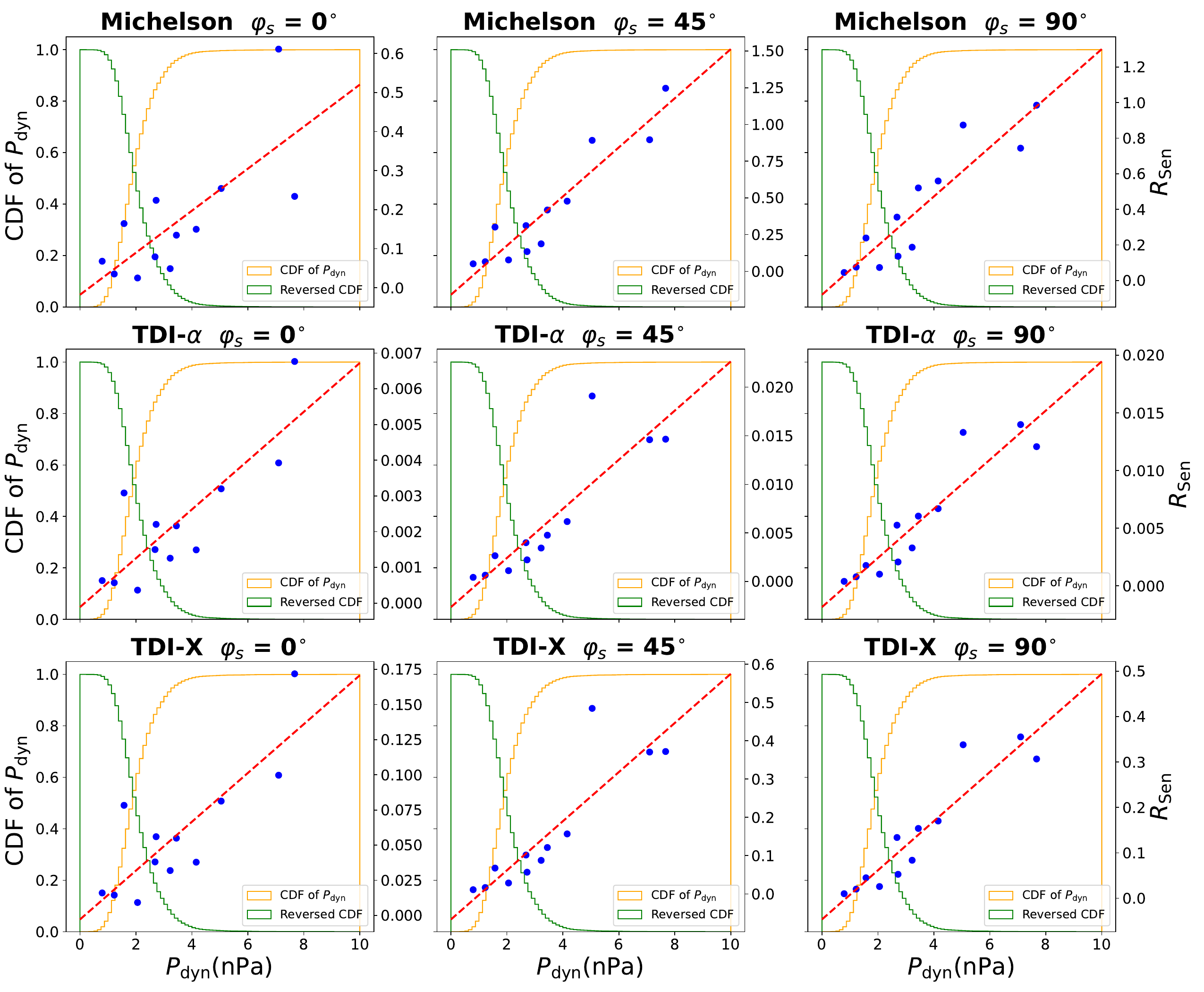}
\caption{Relationship between $R_{\mathrm{sen}}$ and $\overline{P}_{\rm dyn}$ at 10 mHz, and statistical results of $\overline{P}_{\rm dyn}$. The blue points represent the values of $R_{\mathrm{sen}}$ at 10 mHz calculated with different $\overline{P}_{\rm dyn}$, and the red dashed line is the result of the linear fitting. The orange and green lines depict the CDF and reversed CDF of $\overline{P}_{\rm dyn}$ during 1999 -- 2021, respectively.
\label{fig: statistic 10mHz}}
\end{figure}

In order to quantitatively describe the OPD noise, we calculated the ratio of $\Delta l/L$ to the equivalent strain noise requirements for Michelson, TDI-$\alpha$, and TDI-$X$ combinations in the frequency domain. We find the ratio of $\Delta l/L$ to the equivalent strain noise requirements is maximal at around 10 mHz in frequency domain for Michelson and TDI-$X$ combination, which is consistent with the result of \citet{Su2021}. Here, we label the ratio of $\Delta l/L$ to the equivalent strain noise requirements for different interferometry combinations at 10 mHz as $R_{\rm sen}$. 
Figure \ref{fig: statistic 10mHz} shows the relationship between $R_{\mathrm{sen}}$ and $\overline{P}_{\rm dyn}$. The orange and green bins represent the cumulative distribution function (CDF) and the reverse CDF of $\overline{P}_{\rm dyn}$ during a TQ orbital cycle (3.65 days) in 1999--2021. 
$R_{\rm sen}$ for Michelson, TDI-$\alpha$, and TDI-$X$ are marked as blue dots in the top, middle, and bottom panels, respectively.



As shown in Figure \ref{fig: statistic 10mHz}, $\overline{P}_{\rm dyn}-R_{\rm sen}$ distributions show the linear relationship between $\overline{P}_{\rm dyn}$ and $R_{\rm sen}$. The red dashed lines in Figure \ref{fig: statistic 10mHz} are the linear fitted results. 
We calculate the linear correlation coefficients between $R_{\mathrm{sen}}$ and $\overline{P}_{\rm dyn}$ for the Michelson combination, the values of the orbital planes of $ \varphi_s$=$0^\circ$, {45}$^\circ$, and {90}$^\circ$ are 0.735, 0.939, and 0.914, respectively. For the TDI-$\alpha$ combination, the linear correlation coefficients are 0.829, 0.875, and 0.930, respectively. And for the TDI-X combination, the linear correlation coefficients are 0.829, 0.875, and 0.930, respectively. The average of linear correlation coefficients is larger than 0.8 with the minimum larger than 0.7, which suggests an obvious linear correlation between $R_{\mathrm{sen}}$ and $\overline{P}_{\rm dyn}$ for all the interferometry combinations and all the orbital planes.

With the linear fitting between $\overline{P}_{\rm dyn}$ and $R_{\mathrm{sen}}$, we can estimate the occurrence probability rate during a whole solar cycle. 
For the Michelson combination, the occurrence rate of $R_{\mathrm{sen}}>1$ for $\varphi_s$=$0^\circ$, {45}$^\circ$, {90}$^\circ$ orbital planes are 0\%, 0.063\%, 0.039\%, respectively. And for TDI-$\alpha$ and TDI-$X$ combinations, $R_{\mathrm{sen}}>1$ never occurs. Focus on $R_{\mathrm{sen}}>1$ of Michelson, TDI-$\alpha$, and TDI-$X$ combinations, it shows that TDI-$\alpha$ can suppress OPD noise by nearly 2 orders of magnitude of Michelson combination, and TDI-$X$ can suppress OPD noise to about 1/3 of that of Michelson combination. 
The laser propagation noise budget is roughly set as 30\% of the overall displacement noise requirement. 
For the Michelson combinations, the occurrence probabilities of $R_{\mathrm{sen}}>0.3$ for the orbital planes $\varphi_s$=$0^\circ$, {45}$^\circ$, {90}$^\circ$ are 0.19\%, 15.62\%, and 15.50\%, respectively. For TDI-$\alpha$ combinations, $R_{\mathrm{sen}}>0.3$ never occurs. For the TDI-$X$ combinations, the occurrence probabilities of $R_{\mathrm{sen}}>0.3$ for the orbital planes $\varphi_s$=$0^\circ$, {45}$^\circ$, {90}$^\circ$ are 0\%, 0.22\%, and 0.10\%, respectively. 
It shows that after using TDI, the laser propagation noise is negligible in most situations.



\section{Conclusion} \label{sec:conclusion}

In this work, we study the laser propagation noise during an extreme solar eruptive event. For the Michelson combination, the ASD of the OPD noise at 1 mHz $\varphi_s=0^{\circ}$, $45^{\circ}$, and $90^{\circ}$ are 2.844 pm, 7.739 pm, and 6.970 pm, respectively. The result is much larger than the result of moderate solar activity condition \citep{Su2021}, and is close to the TQ's displacement measurement requirement.
In addition, we study laser propagation noise for the TDI combinations during an extremely strong solar eruption. For TDI-$\alpha$ combination, the OPD noise is nearly two orders of magnitude below TQ's displacement measurement requirement. For the TDI-$X$ combination, and the OPD noise is about 30\% of TQ's displacement measurement requirement at about 10 mHz, which is about 1/3 of the result for Michelson combination.
It suggests that while the TDI combinations can suppress the OPD noise, the OPD noise still needs to be considered in the cases of strong solar eruptions.

In order to obtain the OPD noise during the whole solar activity cycle, we calculate the OPD noise of 12 cases with different $\overline{P}_{\rm dyn}$.
We investigate the correlation between the OPD noise and space weather parameters. We find that $\overline{P}_{\rm dyn}$, Sym-H index, and $D$st index are significantly correlated with the amplitude of the ASDs for OPD noise, where the linear correlation between $\overline{P}_{\rm dyn}$ and the amplitude of the ASDs is strongest with the correlation coefficients about 0.9.
We find an obvious linear correlation relationship between the amplitude of the ASDs at 1 mHz and $\overline{P}_{\rm dyn}$, and the correlation coefficients are larger than 0.8 for all of the cases.
It suggests that $\overline{P}_{\rm dyn}$ can serve as a valuable index for the evaluation of the laser propagation noise during the whole solar activity cycle.
Then we evaluate the occurrence rate by combining the linear correlation between $\overline{P}_{\rm dyn}$ and laser propagation noise with the CDF distribution of $\overline{P}_{\rm dyn}$ during 1999 -- 2021 (about 2 solar activity cycle).
For the Michelson combination, 
the occurrence rate of the OPD noise exceeding 30\% of TQ's requirement are 0.188\%, 15.62\%, and 15.50\% for $\varphi_s$=$0^\circ$, {45}$^\circ$, {90}$^\circ$ orbital planes, respectively.
It indicates that the OPD noise needs to be taken into account when $\overline{P}_{\rm dyn}$ is strong for the Michelson combination.
For TDI-$\alpha$ combinations, The laser propagation noise never exceeds 30\% of TQ's requirement.
For the TDI-$X$ combinations, the occurrence rate of the laser propagation noise exceeding 30\% of TQ's requirement for $\varphi_s$=$0^\circ$, {45}$^\circ$, {90}$^\circ$ are 0\%, 0.22\%, and 0.102\%, respectively.
It indicates that TDI combinations can significantly reduce the occurrence rate of the OPD noise exceeding TQ's requirement.

In summary, during a strong solar eruption, the impact of laser propagation noise on TQ GW detection needs to be considered, even after using TDI; However, from a statistical perspective, the impact of laser propagation noise on TQ is minimal.


\section*{Acknowledgments}
The simulations used in this work are provided by the Community Coordinated Modeling Center (CCMC) at Goddard Space Flight Center through the request system. 
This work is carried out using the SWMF and BATSRUS tools developed at the University of Michigan’s Center for Space Environment Modeling.
The OMNI data are available at the SPDF‐OMNIWeb Service‐Nasa website (https://omniweb.gsfc.nasa.gov/).
S. W. is supported by the National Key R\&D Program of China (No. 2020YFC2201200), NSFC (Grant No. 122261131504), and Guangdong Provincial Demonstration Center project of Experimental  Education in 2023 (The Center of  Experimental Physics and Astronomy Education).
X. Z. is supported by the National Key R\&D Program of China (Grant Nos. 2022YFC2204600 and 2020YFC2201202) and NSFC (Grant No. 12373116), and Fundamental Research Funds for the Central Universities, Sun Yat-sen University (Grant No. 23xkjc001). 

\appendix

The details of MHD simulations can be find in simulation result of Community Coordinated Modeling Center (CCMC) website. The run ID, time range of the simulation, input data, and $\overline{P}_{\rm dyn}$ for the 12 cases in this study are listed in the following table.

\begin{table}[H]
\caption{Run details of the MHD simulations. }
\centering
\resizebox{\textwidth}{!}{$
\begin{tabular}{l|l|l|l}
\hline
Time range (yyyymmdd/hh:mm~--~mmdd/hh:mm)     & Run ID        &Input data         & $\overline{P}_{\rm dyn}$ (\rm nPa)     \\ \hline
20001126/06:00~--~1130/06:00                  & yanan\_liu\_030823\_1   & OMNI  & 7.67                  \\\hline
20031021/20:00~--~1025/20:00                  & yanan\_liu\_120122\_2   & OMNI  & 5.05                  \\\hline
20031108/06:00~--~1112/06:00                  & yanan\_liu\_120122\_3   & OMNI  & 4.15                  \\\hline
20041211/00:00~--~1215/00:00                  & Yan\_Meng\_123020\_3a   & OMNI  & 3.45                  \\\hline
\multirow{2}{*}{20050114/11:00~--~0118/11:00} & yanan\_liu\_061923\_1   &\multirow{2}{*}{OMNI}  & \multirow{2}{*}{7.10} \\
                                   & yanan\_liu\_061923\_2    &                        \\\hline
20050410/00:00~--~0414/00:00                  & Yan\_Meng\_123120\_2a   & OMNI  & 3.22                  \\\hline
20060618/12:00~--~0622/12:00                  & yanan\_liu\_021023\_2   & OMNI  & 1.22                  \\\hline
20080807/12:00~--~0811/12:00                  & Yan\_Meng\_123120\_4    & ACE   & 2.72                 \\\hline
20100211/16:00~--~0215/16:00                  & yanan\_liu\_030823\_3   & OMNI  & 0.79                  \\\hline
20130514/21:00~--~0518/21:00                  & yanan\_liu\_031523\_1   & OMNI  & 2.68                  \\\hline
20170330/00:00~--~0403/00:00                  & JiXiang\_Zhang\_042923\_1 & OMNI & 2.06                  \\\hline
\multirow{4}{*}{20180517/06:00~--~0521/06:00} & JiaHui\_Peng\_042923\_1   & \multirow{4}{*}{OMNI} & \multirow{4}{*}{1.57} \\
                                   & JiaHui\_Peng\_042923\_2   &                        \\
                                   & JiaHui\_Peng\_042923\_3   &                        \\
                                   & JiaHui\_Peng\_042923\_4   &                        \\ \hline
\end{tabular}
$}
\end{table}


\bibliography{sample631}{}

\begin{thebibliography}{}
\expandafter\ifx\csname natexlab\endcsname\relax\def\natexlab#1{#1}\fi
\providecommand{\url}[1]{\href{#1}{#1}}
\providecommand{\dodoi}[1]{doi:~\href{http://doi.org/#1}{\nolinkurl{#1}}}
\providecommand{\doeprint}[1]{\href{http://ascl.net/#1}{\nolinkurl{http://ascl.net/#1}}}
\providecommand{\doarXiv}[1]{\href{https://arxiv.org/abs/#1}{\nolinkurl{https://arxiv.org/abs/#1}}}

\bibitem[{{Abbott} {et~al.}(2016){Abbott}, {Abbott}, {Abbott}, {Abernathy}, {Acernese}, {Ackley}, {Adams}, {Adams}, {Addesso}, {Adhikari}, {Adya}, {Affeldt}, {Agathos}, {Agatsuma}, {Aggarwal}, {Aguiar}, {Aiello}, {Ain}, {Ajith}, {Allen}, {Allocca}, {Altin}, {Anderson}, {Anderson}, {Arai}, {Arain}, {Araya}, {Arceneaux}, {Areeda}, {Arnaud}, {Arun}, {Ascenzi}, {Ashton}, {Ast}, {Aston}, {Astone}, {Aufmuth}, {Aulbert}, {Babak}, {Bacon}, {Bader}, {Baker}, {Baldaccini}, {Ballardin}, {Ballmer}, {Barayoga}, {Barclay}, {Barish}, {Barker}, {Barone}, {Barr}, {Barsotti}, {Barsuglia}, {Barta}, {Bartlett}, {Barton}, {Bartos}, {Bassiri}, {Basti}, {Batch}, {Baune}, {Bavigadda}, {Bazzan}, {Behnke}, {Bejger}, {Belczynski}, {Bell}, {Bell}, {Berger}, {Bergman}, {Bergmann}, {Berry}, {Bersanetti}, {Bertolini}, {Betzwieser}, {Bhagwat}, {Bhandare}, {Bilenko}, {Billingsley}, {Birch}, {Birney}, {Birnholtz}, {Biscans}, {Bisht}, {Bitossi}, {Biwer}, {Bizouard}, {Blackburn}, {Blair}, {Blair}, {Blair}, {Bloemen}, {Bock}, {Bodiya}, {Boer},
  {Bogaert}, {Bogan}, {Bohe}, {Bojtos}, {Bond}, {Bondu}, {Bonnand}, {Boom}, {Bork}, {Boschi}, {Bose}, {Bouffanais}, {Bozzi}, {Bradaschia}, {Brady}, {Braginsky}, {Branchesi}, {Brau}, {Briant}, {Brillet}, {Brinkmann}, {Brisson}, {Brockill}, {Brooks}, {Brown}, {Brown}, {Brown}, {Buchanan}, {Buikema}, {Bulik}, {Bulten}, {Buonanno}, {Buskulic}, {Buy}, {Byer}, {Cabero}, {Cadonati}, {Cagnoli}, {Cahillane}, {Bustillo}, {Callister}, {Calloni}, {Camp}, {Cannon}, {Cao}, {Capano}, {Capocasa}, {Carbognani}, {Caride}, {Casanueva Diaz}, {Casentini}, {Caudill}, {Cavagli{\`a}}, {Cavalier}, {Cavalieri}, {Cella}, {Cepeda}, {Baiardi}, {Cerretani}, {Cesarini}, {Chakraborty}, {Chalermsongsak}, {Chamberlin}, {Chan}, {Chao}, {Charlton}, {Chassande-Mottin}, {Chen}, {Chen}, {Cheng}, {Chincarini}, {Chiummo}, {Cho}, {Cho}, {Chow}, {Christensen}, {Chu}, {Chua}, {Chung}, {Ciani}, {Clara}, {Clark}, {Cleva}, {Coccia}, {Cohadon}, {Colla}, {Collette}, {Cominsky}, {Constancio}, {Conte}, {Conti}, {Cook}, {Corbitt}, {Cornish}, {Corsi},
  {Cortese}, {Costa}, {Coughlin}, {Coughlin}, {Coulon}, {Countryman}, {Couvares}, {Cowan}, {Coward}, {Cowart}, {Coyne}, {Coyne}, {Craig}, {Creighton}, {Creighton}, {Cripe}, {Crowder}, {Cruise}, {Cumming}, {Cunningham}, {Cuoco}, {Dal Canton}, {Danilishin}, {D'Antonio}, {Danzmann}, {Darman}, {Da Silva Costa}, {Dattilo}, {Dave}, {Daveloza}, {Davier}, {Davies}, {Daw}, {Day}, {De}, {DeBra}, {Debreczeni}, {Degallaix}, {De Laurentis}, {Del{\'e}glise}, {Del Pozzo}, {Denker}, {Dent}, {Dereli}, {Dergachev}, {DeRosa}, {De Rosa}, {DeSalvo}, {Dhurandhar}, {D{\'\i}az}, {Di Fiore}, {Di Giovanni}, {Di Lieto}, {Di Pace}, {Di Palma}, {Di Virgilio}, {Dojcinoski}, {Dolique}, {Donovan}, {Dooley}, {Doravari}, {Douglas}, {Downes}, {Drago}, {Drever}, {Driggers}, {Du}, {Ducrot}, {Dwyer}, {Edo}, {Edwards}, {Effler}, {Eggenstein}, {Ehrens}, {Eichholz}, {Eikenberry}, {Engels}, {Essick}, {Etzel}, {Evans}, {Evans}, {Everett}, {Factourovich}, {Fafone}, {Fair}, {Fairhurst}, {Fan}, {Fang}, {Farinon}, {Farr}, {Farr}, {Favata}, {Fays},
  {Fehrmann}, {Fejer}, {Feldbaum}, {Ferrante}, {Ferreira}, {Ferrini}, {Fidecaro}, {Finn}, {Fiori}, {Fiorucci}, {Fisher}, {Flaminio}, {Fletcher}, {Fong}, {Fournier}, {Franco}, {Frasca}, {Frasconi}, {Frede}, {Frei}, {Freise}, {Frey}, {Frey}, {Fricke}, {Fritschel}, {Frolov}, {Fulda}, {Fyffe}, {Gabbard}, {Gair}, {Gammaitoni}, {Gaonkar}, {Garufi}, {Gatto}, {Gaur}, {Gehrels}, {Gemme}, {Gendre}, {Genin}, {Gennai}, {George}, {Gergely}, {Germain}, {Ghosh}, {Ghosh}, {Ghosh}, {Giaime}, {Giardina}, {Giazotto}, {Gill}, {Glaefke}, {Gleason}, {Goetz}, {Goetz}, {Gondan}, {Gonz{\'a}lez}, {Castro}, {Gopakumar}, {Gordon}, {Gorodetsky}, {Gossan}, {Gosselin}, {Gouaty}, {Graef}, {Graff}, {Granata}, {Grant}, {Gras}, {Gray}, {Greco}, {Green}, {Greenhalgh}, {Groot}, {Grote}, {Grunewald}, {Guidi}, {Guo}, {Gupta}, {Gupta}, {Gushwa}, {Gustafson}, {Gustafson}, {Hacker}, {Hall}, {Hall}, {Hammond}, {Haney}, {Hanke}, {Hanks}, {Hanna}, {Hannam}, {Hanson}, {Hardwick}, {Harms}, {Harry}, {Harry}, {Hart}, {Hartman}, {Haster}, {Haughian},
  {Healy}, {Heefner}, {Heidmann}, {Heintze}, {Heinzel}, {Heitmann}, {Hello}, {Hemming}, {Hendry}, {Heng}, {Hennig}, {Heptonstall}, {Heurs}, {Hild}, {Hoak}, {Hodge}, {Hofman}, {Hollitt}, {Holt}, {Holz}, {Hopkins}, {Hosken}, {Hough}, {Houston}, {Howell}, {Hu}, {Huang}, {Huerta}, {Huet}, {Hughey}, {Husa}, {Huttner}, {Huynh-Dinh}, {Idrisy}, {Indik}, {Ingram}, {Inta}, {Isa}, {Isac}, {Isi}, {Islas}, {Isogai}, {Iyer}, {Izumi}, {Jacobson}, {Jacqmin}, {Jang}, {Jani}, {Jaranowski}, {Jawahar}, {Jim{\'e}nez-Forteza}, {Johnson}, {Johnson-McDaniel}, {Jones}, {Jones}, {Jonker}, {Ju}, {Haris}, {Kalaghatgi}, {Kalogera}, {Kandhasamy}, {Kang}, {Kanner}, {Karki}, {Kasprzack}, {Katsavounidis}, {Katzman}, {Kaufer}, {Kaur}, {Kawabe}, {Kawazoe}, {K{\'e}f{\'e}lian}, {Kehl}, {Keitel}, {Kelley}, {Kells}, {Kennedy}, {Keppel}, {Key}, {Khalaidovski}, {Khalili}, {Khan}, {Khan}, {Khan}, {Khazanov}, {Kijbunchoo}, {Kim}, {Kim}, {Kim}, {Kim}, {Kim}, {Kim}, {King}, {King}, {Kinzel}, {Kissel}, {Kleybolte}, {Klimenko}, {Koehlenbeck}, {Kokeyama},
  {Koley}, {Kondrashov}, {Kontos}, {Koranda}, {Korobko}, {Korth}, {Kowalska}, {Kozak}, {Kringel}, {Krishnan}, {Kr{\'o}lak}, {Krueger}, {Kuehn}, {Kumar}, {Kumar}, {Kuo}, {Kutynia}, {Kwee}, {Lackey}, {Landry}, {Lange}, {Lantz}, {Lasky}, {Lazzarini}, {Lazzaro}, {Leaci}, {Leavey}, {Lebigot}, {Lee}, {Lee}, {Lee}, {Lee}, {Lenon}, {Leonardi}, {Leong}, {Leroy}, {Letendre}, {Levin}, {Levine}, {Li}, {Libson}, {Littenberg}, {Lockerbie}, {Logue}, {Lombardi}, {London}, {Lord}, {Lorenzini}, {Loriette}, {Lormand}, {Losurdo}, {Lough}, {Lousto}, {Lovelace}, {L{\"u}ck}, {Lundgren}, {Luo}, {Lynch}, {Ma}, {MacDonald}, {Machenschalk}, {MacInnis}, {Macleod}, {Maga{\~n}a-Sandoval}, {Magee}, {Mageswaran}, {Majorana}, {Maksimovic}, {Malvezzi}, {Man}, {Mandel}, {Mandic}, {Mangano}, {Mansell}, {Manske}, {Mantovani}, {Marchesoni}, {Marion}, {M{\'a}rka}, {M{\'a}rka}, {Markosyan}, {Maros}, {Martelli}, {Martellini}, {Martin}, {Martin}, {Martynov}, {Marx}, {Mason}, {Masserot}, {Massinger}, {Masso-Reid}, {Matichard}, {Matone}, {Mavalvala},
  {Mazumder}, {Mazzolo}, {McCarthy}, {McClelland}, {McCormick}, {McGuire}, {McIntyre}, {McIver}, {McManus}, {McWilliams}, {Meacher}, {Meadors}, {Meidam}, {Melatos}, {Mendell}, {Mendoza-Gandara}, {Mercer}, {Merilh}, {Merzougui}, {Meshkov}, {Messenger}, {Messick}, {Meyers}, {Mezzani}, {Miao}, {Michel}, {Middleton}, {Mikhailov}, {Milano}, {Miller}, {Millhouse}, {Minenkov}, {Ming}, {Mirshekari}, {Mishra}, {Mitra}, {Mitrofanov}, {Mitselmakher}, {Mittleman}, {Moggi}, {Mohan}, {Mohapatra}, {Montani}, {Moore}, {Moore}, {Moraru}, {Moreno}, {Morriss}, {Mossavi}, {Mours}, {Mow-Lowry}, {Mueller}, {Mueller}, {Muir}, {Mukherjee}, {Mukherjee}, {Mukherjee}, {Mukund}, {Mullavey}, {Munch}, {Murphy}, {Murray}, {Mytidis}, {Nardecchia}, {Naticchioni}, {Nayak}, {Necula}, {Nedkova}, {Nelemans}, {Neri}, {Neunzert}, {Newton}, {Nguyen}, {Nielsen}, {Nissanke}, {Nitz}, {Nocera}, {Nolting}, {Normandin}, {Nuttall}, {Oberling}, {Ochsner}, {O'Dell}, {Oelker}, {Ogin}, {Oh}, {Oh}, {Ohme}, {Oliver}, {Oppermann}, {Oram}, {O'Reilly},
  {O'Shaughnessy}, {Ott}, {Ottaway}, {Ottens}, {Overmier}, {Owen}, {Pai}, {Pai}, {Palamos}, {Palashov}, {Palomba}, {Pal-Singh}, {Pan}, {Pan}, {Pankow}, {Pannarale}, {Pant}, {Paoletti}, {Paoli}, {Papa}, {Paris}, {Parker}, {Pascucci}, {Pasqualetti}, {Passaquieti}, {Passuello}, {Patricelli}, {Patrick}, {Pearlstone}, {Pedraza}, {Pedurand}, {Pekowsky}, {Pele}, {Penn}, {Perreca}, {Pfeiffer}, {Phelps}, {Piccinni}, {Pichot}, {Pickenpack}, {Piergiovanni}, {Pierro}, {Pillant}, {Pinard}, {Pinto}, {Pitkin}, {Poeld}, {Poggiani}, {Popolizio}, {Post}, {Powell}, {Prasad}, {Predoi}, {Premachandra}, {Prestegard}, {Price}, {Prijatelj}, {Principe}, {Privitera}, {Prix}, {Prodi}, {Prokhorov}, {Puncken}, {Punturo}, {Puppo}, {P{\"u}rrer}, {Qi}, {Qin}, {Quetschke}, {Quintero}, {Quitzow-James}, {Raab}, {Rabeling}, {Radkins}, {Raffai}, {Raja}, {Rakhmanov}, {Ramet}, {Rapagnani}, {Raymond}, {Razzano}, {Re}, {Read}, {Reed}, {Regimbau}, {Rei}, {Reid}, {Reitze}, {Rew}, {Reyes}, {Ricci}, {Riles}, {Robertson}, {Robie}, {Robinet}, {Rocchi},
  {Rolland}, {Rollins}, {Roma}, {Romano}, {Romano}, {Romanov}, {Romie}, {Rosi{\'n}ska}, {Rowan}, {R{\"u}diger}, {Ruggi}, {Ryan}, {Sachdev}, {Sadecki}, {Sadeghian}, {Salconi}, {Saleem}, {Salemi}, {Samajdar}, {Sammut}, {Sampson}, {Sanchez}, {Sandberg}, {Sandeen}, {Sanders}, {Sanders}, {Sassolas}, {Sathyaprakash}, {Saulson}, {Sauter}, {Savage}, {Sawadsky}, {Schale}, {Schilling}, {Schmidt}, {Schmidt}, {Schnabel}, {Schofield}, {Sch{\"o}nbeck}, {Schreiber}, {Schuette}, {Schutz}, {Scott}, {Scott}, {Sellers}, {Sengupta}, {Sentenac}, {Sequino}, {Sergeev}, {Serna}, {Setyawati}, {Sevigny}, {Shaddock}, {Shaffer}, {Shah}, {Shahriar}, {Shaltev}, {Shao}, {Shapiro}, {Shawhan}, {Sheperd}, {Shoemaker}, {Shoemaker}, {Siellez}, {Siemens}, {Sigg}, {Silva}, {Simakov}, {Singer}, {Singer}, {Singh}, {Singh}, {Singhal}, {Sintes}, {Slagmolen}, {Smith}, {Smith}, {Smith}, {Smith}, {Son}, {Sorazu}, {Sorrentino}, {Souradeep}, {Srivastava}, {Staley}, {Steinke}, {Steinlechner}, {Steinlechner}, {Steinmeyer}, {Stephens}, {Stevenson}, {Stone},
  {Strain}, {Straniero}, {Stratta}, {Strauss}, {Strigin}, {Sturani}, {Stuver}, {Summerscales}, {Sun}, {Sutton}, {Swinkels}, {Szczepa{\'n}czyk}, {Tacca}, {Talukder}, {Tanner}, {T{\'a}pai}, {Tarabrin}, {Taracchini}, {Taylor}, {Theeg}, {Thirugnanasambandam}, {Thomas}, {Thomas}, {Thomas}, {Thorne}, {Thorne}, {Thrane}, {Tiwari}, {Tiwari}, {Tokmakov}, {Tomlinson}, {Tonelli}, {Torres}, {Torrie}, {T{\"o}yr{\"a}}, {Travasso}, {Traylor}, {Trifir{\`o}}, {Tringali}, {Trozzo}, {Tse}, {Turconi}, {Tuyenbayev}, {Ugolini}, {Unnikrishnan}, {Urban}, {Usman}, {Vahlbruch}, {Vajente}, {Valdes}, {Vallisneri}, {van Bakel}, {van Beuzekom}, {van den Brand}, {Van Den Broeck}, {Vander-Hyde}, {van der Schaaf}, {van Heijningen}, {van Veggel}, {Vardaro}, {Vass}, {Vas{\'u}th}, {Vaulin}, {Vecchio}, {Vedovato}, {Veitch}, {Veitch}, {Venkateswara}, {Verkindt}, {Vetrano}, {Vicer{\'e}}, {Vinciguerra}, {Vine}, {Vinet}, {Vitale}, {Vo}, {Vocca}, {Vorvick}, {Voss}, {Vousden}, {Vyatchanin}, {Wade}, {Wade}, {Wade}, {Waldman}, {Walker}, {Wallace},
  {Walsh}, {Wang}, {Wang}, {Wang}, {Wang}, {Wang}, {Ward}, {Ward}, {Warner}, {Was}, {Weaver}, {Wei}, {Weinert}, {Weinstein}, {Weiss}, {Welborn}, {Wen}, {We{\ss}els}, {Westphal}, {Wette}, {Whelan}, {Whitcomb}, {White}, {Whiting}, {Wiesner}, {Wilkinson}, {Willems}, {Williams}, {Williams}, {Williamson}, {Willis}, {Willke}, {Wimmer}, {Winkelmann}, {Winkler}, {Wipf}, {Wiseman}, {Wittel}, {Woan}, {Worden}, {Wright}, {Wu}, {Yablon}, {Yakushin}, {Yam}, {Yamamoto}, {Yancey}, {Yap}, {Yu}, {Yvert}, {Zadro{\.Z}ny}, {Zangrando}, {Zanolin}, {Zendri}, {Zevin}, {Zhang}, {Zhang}, {Zhang}, {Zhang}, {Zhao}, {Zhou}, {Zhou}, {Zhu}, {Zucker}, {Zuraw}, {Zweizig}, {LIGO Scientific Collaboration}, \& {Virgo Collaboration}}]{Abbott2016}
{Abbott}, B.~P., {Abbott}, R., {Abbott}, T.~D., {et~al.} 2016, \prl, 116, 061102, \dodoi{10.1103/PhysRevLett.116.061102}

\bibitem[{{Amaro-Seoane} {et~al.}(2017){Amaro-Seoane}, {Audley}, {Babak}, {Baker}, {Barausse}, {Bender}, {Berti}, {Binetruy}, {Born}, {Bortoluzzi}, {Camp}, {Caprini}, {Cardoso}, {Colpi}, {Conklin}, {Cornish}, {Cutler}, {Danzmann}, {Dolesi}, {Ferraioli}, {Ferroni}, {Fitzsimons}, {Gair}, {Gesa Bote}, {Giardini}, {Gibert}, {Grimani}, {Halloin}, {Heinzel}, {Hertog}, {Hewitson}, {Holley-Bockelmann}, {Hollington}, {Hueller}, {Inchauspe}, {Jetzer}, {Karnesis}, {Killow}, {Klein}, {Klipstein}, {Korsakova}, {Larson}, {Livas}, {Lloro}, {Man}, {Mance}, {Martino}, {Mateos}, {McKenzie}, {McWilliams}, {Miller}, {Mueller}, {Nardini}, {Nelemans}, {Nofrarias}, {Petiteau}, {Pivato}, {Plagnol}, {Porter}, {Reiche}, {Robertson}, {Robertson}, {Rossi}, {Russano}, {Schutz}, {Sesana}, {Shoemaker}, {Slutsky}, {Sopuerta}, {Sumner}, {Tamanini}, {Thorpe}, {Troebs}, {Vallisneri}, {Vecchio}, {Vetrugno}, {Vitale}, {Volonteri}, {Wanner}, {Ward}, {Wass}, {Weber}, {Ziemer}, \& {Zweifel}}]{Amaro-Seoane2017}
{Amaro-Seoane}, P., {Audley}, H., {Babak}, S., {et~al.} 2017, arXiv e-prints, arXiv:1702.00786, \dodoi{10.48550/arXiv.1702.00786}

\bibitem[{{Armano} {et~al.}(2020){Armano}, {Audley}, {Baird}, {Binetruy}, {Born}, {Bortoluzzi}, {Castelli}, {Cavalleri}, {Cesarini}, {Cruise}, {Danzmann}, {de Deus Silva}, {Diepholz}, {Dixon}, {Dolesi}, {Ferraioli}, {Ferroni}, {Fitzsimons}, {Freschi}, {Gesa}, {Gibert}, {Giardini}, {Giusteri}, {Grimani}, {Grzymisch}, {Harrison}, {Hartig}, {Heinzel}, {Hewitson}, {Hollington}, {Hoyland}, {Hueller}, {Inchausp{\'e}}, {Jennrich}, {Jetzer}, {Karnesis}, {Kaune}, {Korsakova}, {Killow}, {Lobo}, {Liu}, {L{\'o}pez-Zaragoza}, {Maarschalkerweerd}, {Mance}, {Mart{\'\i}n}, {Martin-Polo}, {Martino}, {Martin-Porqueras}, {Mateos}, {McNamara}, {Mendes}, {Mendes}, {Meshksar}, {Nofrarias}, {Paczkowski}, {Perreur-Lloyd}, {Petiteau}, {Pivato}, {Plagnol}, {Ramos-Castro}, {Reiche}, {Rivas}, {Robertson}, {Roma-Dollase}, {Russano}, {Slutsky}, {Sopuerta}, {Sumner}, {Telloni}, {Texier}, {Thorpe}, {Trenkel}, {Vetrugno}, {Vitale}, {Wanner}, {Ward}, {Wass}, {Wealthy}, {Weber}, {Wissel}, {Wittchen}, \& {Zweifel}}]{Armano2020}
{Armano}, M., {Audley}, H., {Baird}, J., {et~al.} 2020, \mnras, 494, 3014, \dodoi{10.1093/mnras/staa830}

\bibitem[{{Armstrong} {et~al.}(1999){Armstrong}, {Estabrook}, \& {Tinto}}]{1999ApJ...527..814A}
{Armstrong}, J.~W., {Estabrook}, F.~B., \& {Tinto}, M. 1999, \apj, 527, 814, \dodoi{10.1086/308110}

\bibitem[{Burt \& Smith(2012)}]{DSCOVR2012}
Burt, J., \& Smith, B. 2012, in 2012 IEEE Aerospace Conference, 1--13, \dodoi{10.1109/AERO.2012.6187025}

\bibitem[{{Chen}(2011)}]{Chen2011}
{Chen}, P.~F. 2011, Living Reviews in Solar Physics, 8, 1, \dodoi{10.12942/lrsp-2011-1}

\bibitem[{{Chen} {et~al.}(2007){Chen}, {Hu}, \& {Sun}}]{ChenY_2007}
{Chen}, Y., {Hu}, Y.~Q., \& {Sun}, S.~J. 2007, The Astrophysical Journal, 665, 1421, \dodoi{10.1086/519551}

\bibitem[{{Chen} {et~al.}(2022){Chen}, {Liao}, {Li}, {Wang}, {Deng}, \& {Hong}}]{Chen_zhou2022}
{Chen}, Z., {Liao}, W., {Li}, H., {et~al.} 2022, Space Weather, 20, e2021SW002854, \dodoi{10.1029/2021SW00285410.1002/essoar.10507605.1}

\bibitem[{{Gombosi} {et~al.}(2021){Gombosi}, {Chen}, {Glocer}, {Huang}, {Jia}, {Liemohn}, {Manchester}, {Pulkkinen}, {Sachdeva}, {Al Shidi}, {Sokolov}, {Szente}, {Tenishev}, {Toth}, {van der Holst}, {Welling}, {Zhao}, \& {Zou}}]{Gombosi2021}
{Gombosi}, T.~I., {Chen}, Y., {Glocer}, A., {et~al.} 2021, Journal of Space Weather and Space Climate, 11, 42, \dodoi{10.1051/swsc/2021020}

\bibitem[{{Han} {et~al.}(2023){Han}, {Cai}, {Yang}, {Xu}, {Xia}, {Jia}, {Gao}, \& {Han}}]{Han2023}
{Han}, R., {Cai}, M., {Yang}, T., {et~al.} 2023, Scientific Reports, 13, 9932, \dodoi{10.1038/s41598-023-37005-3}

\bibitem[{{Han} {et~al.}(2024){Han}, {Cai}, {Yang}, {Xu}, {Xia}, {Jia}, {Gao}, {Li}, {Zhang}, {Li}, \& {Han}}]{Han2024}
---. 2024, Space Weather, 22, e2023SW003724, \dodoi{10.1029/2023SW003724}

\bibitem[{{Hao} {et~al.}(2015){Hao}, {Fang}, {Cao}, \& {Chen}}]{Hao2015}
{Hao}, Q., {Fang}, C., {Cao}, W., \& {Chen}, P.~F. 2015, \apjs, 221, 33, \dodoi{10.1088/0067-0049/221/2/33}

\bibitem[{{Hartley} {et~al.}(2014){Hartley}, {Denton}, \& {Rodriguez}}]{Hartley2014}
{Hartley}, D.~P., {Denton}, M.~H., \& {Rodriguez}, J.~V. 2014, Journal of Geophysical Research (Space Physics), 119, 4556, \dodoi{10.1002/2014JA019779}

\bibitem[{{Hathaway}(2015)}]{Hathaway2015}
{Hathaway}, D.~H. 2015, Living Reviews in Solar Physics, 12, 4, \dodoi{10.1007/lrsp-2015-4}

\bibitem[{Hu \& Wu(2017)}]{hu2017taiji}
Hu, W.-R., \& Wu, Y.-L. 2017, 4, 685, \dodoi{10.1093/nsr/nwx116}

\bibitem[{{Hu} {et~al.}(2018){Hu}, {Li}, {Wang}, {Feng}, {Zhou}, {Hu}, {Hu}, {Mei}, \& {Shao}}]{2018CQGra..35i5008H}
{Hu}, X.-C., {Li}, X.-H., {Wang}, Y., {et~al.} 2018, Classical and Quantum Gravity, 35, 095008, \dodoi{10.1088/1361-6382/aab52f}

\bibitem[{{Huang} {et~al.}(2021){Huang}, {Xiong}, {Song}, {Nan}, {Yuan}, {Jiang}, {Deng}, \& {Yu}}]{HuangSY2021}
{Huang}, S.~Y., {Xiong}, Q.~Y., {Song}, L.~F., {et~al.} 2021, \apj, 922, 54, \dodoi{10.3847/1538-4357/ac2668}

\bibitem[{{Hutchinson}(2002)}]{2002ppd..book.....H}
{Hutchinson}, I.~H. 2002, {Principles of Plasma Diagnostics}

\bibitem[{{Jennrich} {et~al.}(2021){Jennrich}, {Luetzgendorf}, {Thorpe}, {Slutsky}, \& {Cutler}}]{Jennrich2021}
{Jennrich}, O., {Luetzgendorf}, N., {Thorpe}, J.~I., {Slutsky}, J., \& {Cutler}, C. 2021, \prd, 104, 062003, \dodoi{10.1103/PhysRevD.104.062003}

\bibitem[{{Jia} {et~al.}(2023){Jia}, {Li}, {Qiao}, \& {Zhou}}]{Jia2023}
{Jia}, F., {Li}, X., {Qiao}, D., \& {Zhou}, X. 2023, Acta Astronautica, 202, 522, \dodoi{10.1016/j.actaastro.2022.10.048}

\bibitem[{{Jing} {et~al.}(2022){Jing}, {Zheng}, {Yang}, {Zhang}, {Lu}, {Tang}, \& {Su}}]{Jing2022J}
{Jing}, Y.-D., {Zheng}, L., {Yang}, S., {et~al.} 2022, \prd, 106, 082006, \dodoi{10.1103/PhysRevD.106.082006}

\bibitem[{{Kallenrode}(2004)}]{Kallenrode2004}
{Kallenrode}, M.-B. 2004, {Space Physics : An Introduction to Plasmas and Particles in the Heliosphere and Magnetospheres}, \dodoi{10.1007/978-3-662-09959-9}

\bibitem[{{Kawamura} {et~al.}(2006){Kawamura}, {Nakamura}, {Ando}, {Seto}, {Tsubono}, {Numata}, {Takahashi}, {Nagano}, {Ishikawa}, {Musha}, {Ueda}, {Sato}, {Hosokawa}, {Agatsuma}, {Akutsu}, {Aoyanagi}, {Arai}, {Araya}, {Asada}, {Aso}, {Chiba}, {Ebisuzaki}, {Eriguchi}, {Fujimoto}, {Fukushima}, {Futamase}, {Ganzu}, {Harada}, {Hashimoto}, {Hayama}, {Hikida}, {Himemoto}, {Hirabayashi}, {Hiramatsu}, {Ichiki}, {Ikegami}, {Inoue}, {Ioka}, {Ishidoshiro}, {Itoh}, {Kamagasako}, {Kanda}, {Kawashima}, {Kirihara}, {Kiuchi}, {Kobayashi}, {Kohri}, {Kojima}, {Kokeyama}, {Kozai}, {Kudoh}, {Kunimori}, {Kuroda}, {Maeda}, {Matsuhara}, {Mino}, {Miyakawa}, {Miyoki}, {Mizusawa}, {Morisawa}, {Mukohyama}, {Naito}, {Nakagawa}, {Nakamura}, {Nakano}, {Nakao}, {Nishizawa}, {Niwa}, {Nozawa}, {Ohashi}, {Ohishi}, {Ohkawa}, {Okutomi}, {Oohara}, {Sago}, {Saijo}, {Sakagami}, {Sakata}, {Sasaki}, {Sato}, {Shibata}, {Shinkai}, {Somiya}, {Sotani}, {Sugiyama}, {Tagoshi}, {Takahashi}, {Takahashi}, {Takahashi}, {Takano}, {Tanaka}, {Taniguchi},
  {Taruya}, {Tashiro}, {Tokunari}, {Tsujikawa}, {Tsunesada}, {Yamamoto}, {Yamazaki}, {Yokoyama}, {Yoo}, {Yoshida}, \& {Yoshino}}]{Kawamura2006DECIGO}
{Kawamura}, S., {Nakamura}, T., {Ando}, M., {et~al.} 2006, Classical and Quantum Gravity, 23, S125, \dodoi{10.1088/0264-9381/23/8/S17}

\bibitem[{{Kilpua} {et~al.}(2015){Kilpua}, {Lumme}, {Andreeova}, {Isavnin}, \& {Koskinen}}]{Kilpua2015}
{Kilpua}, E.~K.~J., {Lumme}, E., {Andreeova}, K., {Isavnin}, A., \& {Koskinen}, H.~E.~J. 2015, Journal of Geophysical Research (Space Physics), 120, 4112, \dodoi{10.1002/2015JA021138}

\bibitem[{{King} \& {Papitashvili}(2005)}]{2005JGRA..110.2104K}
{King}, J.~H., \& {Papitashvili}, N.~E. 2005, Journal of Geophysical Research (Space Physics), 110, A02104, \dodoi{10.1029/2004JA010649}

\bibitem[{{Lei} {et~al.}(2024){Lei}, {Su}, {Hong}, {Li}, {Zhao}, {Chen}, {Chu}, {Li}, {Bai}, \& {Zhou}}]{Lei2024}
{Lei}, C., {Su}, W., {Hong}, W., {et~al.} 2024, Classical and Quantum Gravity, 41, 025001, \dodoi{10.1088/1361-6382/ad105a}

\bibitem[{{Li} {et~al.}(2021){Li}, {Khotyaintsev}, {Tang}, {Graham}, {Norgren}, {Vaivads}, {Andr{\'e}}, {Le}, {Egedal}, {Dokgo}, {Fujimoto}, {He}, {Burch}, {Lindqvist}, {Ergun}, {Torbert}, {Le Contel}, {Gershman}, {Giles}, {Lavraud}, {Fuselier}, {Plaschke}, {Russell}, {Guo}, {Lu}, \& {Wang}}]{LiWY2021}
{Li}, W.~Y., {Khotyaintsev}, Y.~V., {Tang}, B.~B., {et~al.} 2021, \grl, 48, e93164, \dodoi{10.1029/2021GL093164}

\bibitem[{{Lu} {et~al.}(2019){Lu}, {Zhang}, {Wang}, {Gu}, \& {Guan}}]{Lu_J_Y2019}
{Lu}, J., {Zhang}, H., {Wang}, M., {Gu}, C., \& {Guan}, H. 2019, Earth and Planetary Physics, 3, 8, \dodoi{10.26464/epp2019002}

\bibitem[{{Lu} {et~al.}(2013){Lu}, {Jing}, {Liu}, {Kabin}, \& {Jiang}}]{Lu_J_Y2013}
{Lu}, J.~Y., {Jing}, H., {Liu}, Z.~Q., {Kabin}, K., \& {Jiang}, Y. 2013, Journal of Geophysical Research (Space Physics), 118, 2021, \dodoi{10.1002/jgra.50093}

\bibitem[{{Lu} {et~al.}(2021){Lu}, {Su}, {Zhang}, {He}, {Duan}, {Jiang}, \& {Yeh}}]{Lu2021}
{Lu}, L.-F., {Su}, W., {Zhang}, X., {et~al.} 2021, Journal of Geophysical Research (Space Physics), 126, e28579, \dodoi{10.1029/2020JA028579}

\bibitem[{{Luo} {et~al.}(2016){Luo}, {Chen}, {Duan}, {Gong}, {Hu}, {Ji}, {Liu}, {Mei}, {Milyukov}, {Sazhin}, {Shao}, {Toth}, {Tu}, {Wang}, {Wang}, {Yeh}, {Zhan}, {Zhang}, {Zharov}, \& {Zhou}}]{Luo2016TQ}
{Luo}, J., {Chen}, L.-S., {Duan}, H.-Z., {et~al.} 2016, Classical and Quantum Gravity, 33, 035010, \dodoi{10.1088/0264-9381/33/3/035010}

\bibitem[{{Maezawa}(1976)}]{Maezawa1976}
{Maezawa}, K. 1976, \jgr, 81, 2289, \dodoi{10.1029/JA081i013p02289}

\bibitem[{{Manchester} {et~al.}(2008){Manchester}, {Vourlidas}, {T{\'o}th}, {Lugaz}, {Roussev}, {Sokolov}, {Gombosi}, {De Zeeuw}, \& {Opher}}]{2008ApJ...684.1448M}
{Manchester}, Ward~B., I., {Vourlidas}, A., {T{\'o}th}, G., {et~al.} 2008, \apj, 684, 1448, \dodoi{10.1086/590231}

\bibitem[{Muratore(2021)}]{muratore2021time}
Muratore, M. 2021, Doctoral thesis, University of Trento, Trento, Italy, \dodoi{10.15168/11572_312487}

\bibitem[{Ogilvie \& Desch(1997)}]{WIND1997}
Ogilvie, K., \& Desch, M. 1997, Advances in Space Research, 20, 559, \dodoi{https://doi.org/10.1016/S0273-1177(97)00439-0}

\bibitem[{{Oka} {et~al.}(2023){Oka}, {Birn}, {Egedal}, {Guo}, {Ergun}, {Turner}, {Khotyaintsev}, {Hwang}, {Cohen}, \& {Drake}}]{Oka2023}
{Oka}, M., {Birn}, J., {Egedal}, J., {et~al.} 2023, \ssr, 219, 75, \dodoi{10.1007/s11214-023-01011-8}

\bibitem[{Pisacane(2008)}]{pisacane2008space}
Pisacane, V.~L. 2008, The Space Environment and Its Effects on Space Systems (American Institute of Aeronautics and Astronautics), \dodoi{10.2514/4.862533}

\bibitem[{{Qiao} {et~al.}(2023){Qiao}, {Jia}, {Li}, \& {Zhou}}]{Qiao2023}
{Qiao}, D., {Jia}, F., {Li}, X., \& {Zhou}, X. 2023, Space: Science and Technology, 3, 0015, \dodoi{10.34133/space.0015}

\bibitem[{{Rufenach} {et~al.}(1992){Rufenach}, {McPherron}, \& {Schaper}}]{Rufenach1992}
{Rufenach}, C.~L., {McPherron}, R.~L., \& {Schaper}, J. 1992, \jgr, 97, 25, \dodoi{10.1029/91JA02135}

\bibitem[{{Samsonov} {et~al.}(2019){Samsonov}, {Bogdanova}, {Branduardi-Raymont}, {Safrankova}, {Nemecek}, \& {Park}}]{Samsonov2019}
{Samsonov}, A.~A., {Bogdanova}, Y.~V., {Branduardi-Raymont}, G., {et~al.} 2019, Journal of Geophysical Research (Space Physics), 124, 4049, \dodoi{10.1029/2018JA026355}

\bibitem[{{Samsonov} {et~al.}(2016){Samsonov}, {Gordeev}, {Tsyganenko}, {{\r{A}} afr{\'a}nkov{\'a}}, {N{\v{e}}me{\v{c}}ek}, {{\r{A}} im{\r{A}}{\textasciimacron}nek}, {Sibeck}, {T{\'o}th}, {Merkin}, \& {Raeder}}]{Samsonov2016}
{Samsonov}, A.~A., {Gordeev}, E., {Tsyganenko}, N.~A., {et~al.} 2016, Journal of Geophysical Research (Space Physics), 121, 6493, \dodoi{10.1002/2016JA022471}

\bibitem[{{Sharma} \& {Curtis}(2005)}]{Sharma2005}
{Sharma}, A.~S., \& {Curtis}, S.~A. 2005, in Astrophysics and Space Science Library, Vol. 321, Astrophysics and Space Science Library, ed. A.~S. {Sharma} \& P.~K. {Kaw}, 179, \dodoi{10.1007/1-4020-3109-2_8}

\bibitem[{{Shen} {et~al.}(2012{\natexlab{a}}){Shen}, {Liu}, \& {Su}}]{ShenYD2012_1}
{Shen}, Y., {Liu}, Y., \& {Su}, J. 2012{\natexlab{a}}, \apj, 750, 12, \dodoi{10.1088/0004-637X/750/1/12}

\bibitem[{{Shen} {et~al.}(2012{\natexlab{b}}){Shen}, {Liu}, {Su}, \& {Deng}}]{ShenYD2012_2}
{Shen}, Y., {Liu}, Y., {Su}, J., \& {Deng}, Y. 2012{\natexlab{b}}, \apj, 745, 164, \dodoi{10.1088/0004-637X/745/2/164}

\bibitem[{{Smetana}(2020)}]{Smetana2020}
{Smetana}, A. 2020, \mnras, 499, L77, \dodoi{10.1093/mnrasl/slaa155}

\bibitem[{{Smith} {et~al.}(1964){Smith}, {Sonett}, \& {Dungey}}]{MagneticStorms1964}
{Smith}, E.~J., {Sonett}, C.~P., \& {Dungey}, J.~W. 1964, \jgr, 69, 2669, \dodoi{10.1029/JZ069i013p02669}

\bibitem[{{Stone} {et~al.}(1998){Stone}, {Frandsen}, {Mewaldt}, {Christian}, {Margolies}, {Ormes}, \& {Snow}}]{ACE1998}
{Stone}, E.~C., {Frandsen}, A.~M., {Mewaldt}, R.~A., {et~al.} 1998, \ssr, 86, 1, \dodoi{10.1023/A:1005082526237}

\bibitem[{{Su} {et~al.}(2020){Su}, {Wang}, {Zhou}, {Bai}, {Guo}, {Zhou}, {Lee}, {Wang}, {Zhou}, {Shi}, {Yin}, \& {Zhang}}]{Su2020}
{Su}, W., {Wang}, Y., {Zhou}, Z.-B., {et~al.} 2020, Classical and Quantum Gravity, 37, 185017, \dodoi{10.1088/1361-6382/aba181}

\bibitem[{{Su} {et~al.}(2021){Su}, {Wang}, {Zhou}, {Lu}, {Zhou}, {Li}, {Shi}, {Hu}, {Zhou}, {Wang}, {Yeh}, {Wang}, \& {Chen}}]{Su2021}
{Su}, W., {Wang}, Y., {Zhou}, C., {et~al.} 2021, \apj, 914, 139, \dodoi{10.3847/1538-4357/abfc49}

\bibitem[{{Su} {et~al.}(2023){Su}, {Zhou}, {Wang}, {Zhou}, {Chen}, {Hong}, {Peng}, {Yang}, \& {Ni}}]{Su2023}
{Su}, W., {Zhou}, Z.-B., {Wang}, Y., {et~al.} 2023, \prd, 108, 103030, \dodoi{10.1103/PhysRevD.108.103030}

\bibitem[{{Sun} {et~al.}(2023){Sun}, {Li}, {Li}, {yu}, {Feng}, {Li}, {Jin}, \& {Wang}}]{SunZhiyin2023}
{Sun}, Z., {Li}, P., {Li}, Z., {et~al.} 2023, Results in Physics, 53, 106955, \dodoi{10.1016/j.rinp.2023.106955}

\bibitem[{{T{\'o}th} {et~al.}(2005){T{\'o}th}, {Sokolov}, {Gombosi}, {Chesney}, {Clauer}, {de Zeeuw}, {Hansen}, {Kane}, {Manchester}, {Oehmke}, {Powell}, {Ridley}, {Roussev}, {Stout}, {Volberg}, {Wolf}, {Sazykin}, {Chan}, {Yu}, \& {K{\'o}ta}}]{2005JGRA..11012226T}
{T{\'o}th}, G., {Sokolov}, I.~V., {Gombosi}, T.~I., {et~al.} 2005, Journal of Geophysical Research (Space Physics), 110, A12226, \dodoi{10.1029/2005JA011126}

\bibitem[{Veselovsky {et~al.}(2004)Veselovsky, Panasyuk, Avdyushin, Bazilevskaya, Belov, Bogachev, Bogod, Bogomolov, Bothmer, Boyarchuk, Vashenyuk, Vlasov, Gnezdilov, Gorgutsa, Grechnev, Denisov, Dmitriev, Dryer, Yermolaev, Eroshenko, Zherebtsov, Zhitnik, Zhukov, Zastenker, Zelenyi, Zeldovich, Ivanov-Kholodnyi, Ignat'ev, Ishkov, Kolomiytsev, Krasheninnikov, Kudela, Kuzhevsky, Kuzin, Kuznetsov, Kuznetsov, Kurt, Lazutin, Leshchenko, Litvak, Logachev, Lawrence, Markeev, Makhmutov, Mitrofanov, Mitrofanov, Morozov, Myagkova, Nusinov, Oparin, Panasenco, Pertsov, Petrukovich, Podorol'sky, Romashets, Svertilov, Svidsky, Svirzhevskaya, Svirzhevsky, Slemzin, Smith, Sobel'man, Sobolev, Stozhkov, Suvorova, Sukhodrev, Tindo, Tokhchukova, Fomichev, Chashey, Chertok, Shishov, Yushkov, Yakovchouk, \& Yanke}]{veselovsky2004solar}
Veselovsky, I.~S., Panasyuk, M.~I., Avdyushin, S.~I., {et~al.} 2004, 42, 435, \dodoi{10.1023/B:COSM.0000046229.24716.02}

\bibitem[{{Wang} {et~al.}(2009){Wang}, {Liu}, {Li}, {Huang}, {Richardson}, \& {Kan}}]{Wang_C2009}
{Wang}, C., {Liu}, J.~B., {Li}, H., {et~al.} 2009, Journal of Geophysical Research (Space Physics), 114, A05211, \dodoi{10.1029/2008JA013794}

\bibitem[{{Wang} {et~al.}(2015){Wang}, {Lu}, {Yuan}, {Kabin}, {Liu}, {Zhao}, \& {Li}}]{Wang_M2015}
{Wang}, M., {Lu}, J.~Y., {Yuan}, H.~Z., {et~al.} 2015, \planss, 106, 99, \dodoi{10.1016/j.pss.2014.12.002}

\bibitem[{{Ye} {et~al.}(2021){Ye}, {Zhang}, {Ding}, \& {Meng}}]{Ye2021}
{Ye}, B., {Zhang}, X., {Ding}, Y., \& {Meng}, Y. 2021, \prd, 103, 042007, \dodoi{10.1103/PhysRevD.103.042007}

\bibitem[{{Zheng} {et~al.}(2020){Zheng}, {Chen}, \& {Wang}}]{ZhengRuisheng2020}
{Zheng}, R., {Chen}, Y., \& {Wang}, B. 2020, \apjl, 897, L21, \dodoi{10.3847/2041-8213/ab9ebd}

\bibitem[{{Zhou} {et~al.}(2018){Zhou}, {Berchem}, {Walker}, {El-Alaoui}, {Goldstein}, {Lapenta}, {Deng}, {Li}, {Le Contel}, {Graham}, {Lavraud}, {Paterson}, {Giles}, {Burch}, {Torbert}, {Russell}, {Strangeway}, {Zhao}, {Ergun}, {Lindqvist}, \& {Marklund}}]{Zhou2018}
{Zhou}, M., {Berchem}, J., {Walker}, R.~J., {et~al.} 2018, Journal of Geophysical Research (Space Physics), 123, 1834, \dodoi{10.1002/2017JA024517}

\end{thebibliography}
\bibliographystyle{aasjournal}



\end{CJK*}
\end{document}